\documentstyle[12pt,amssymb]{article}
\begin{document}
\tolerance=5000
\def\pp{{\, \mid \hskip -1.5mm =}}
\def\cL{{\cal L}}
\def\be{\begin{equation}}
\def\ee{\end{equation}}
\def\bea{\begin{eqnarray}}
\def\eea{\end{eqnarray}}
\def\tr{{\rm tr}\, }
\def\nn{\nonumber \\}
\def\e{{\rm e}}

\begin{titlepage}

\begin{center}
\Large {\bf Late-time cosmology in (phantom) scalar-tensor theory:
dark energy and the cosmic speed-up}

\vspace*{5mm}

\normalsize

\large{ Emilio Elizalde,$^{\star\heartsuit}$\footnote{Electronic
mail: elizalde@ieec.fcr.es, elizalde@math.mit.edu. Presently on
leave at Department of Mathematics, MIT, 77 Massachusetts Ave,
Cambridge, MA 02139.} Shin'ichi
Nojiri,$^\spadesuit$\footnote{Electronic mail: nojiri@nda.ac.jp,
snojiri@yukawa.kyoto-u.ac.jp} \\
and Sergei D.
Odintsov$^{\clubsuit\heartsuit}$\footnote{Electronic mail:
odintsov@ieec.fcr.es Also at TSPU, Tomsk, Russia}}

\normalsize

\vspace*{3mm}

{\em $\star$ Instituto de Ciencias del Espacio (ICE) \\
Consejo Superior de Investigaciones Cient\'{\i}ficas (CSIC)}
\medskip

{\em $\heartsuit$ Institut d'Estudis Espacials de Catalunya (IEEC), \\
Edifici Nexus, Gran Capit\`a 2-4, 08034 Barcelona, Spain}
\medskip

{\em $\spadesuit$ Department of Applied Physics,
National Defence Academy, \\
Hashirimizu Yokosuka 239-8686, Japan}
\medskip

{\em $\clubsuit$ Instituci\`o Catalana de Recerca i Estudis
Avan\c{c}ats (ICREA), \\Barcelona, Spain}

\end{center}

\vspace*{1mm}

\begin{abstract}

We consider late-time cosmology in a (phantom) scalar-tensor theory
with an exponential potential, as a dark energy model with equation of state
parameter close to -1 (a bit above or below this value).
Scalar (and also other kinds of) matter can be easily taken into account.
An exact spatially-flat
FRW cosmology is constructed for such theory, which admits
(eternal or transient) acceleration phases for the current universe,
in correspondence with observational results.
Some remarks on the possible origin of the phantom, starting from a more
fundamental theory, are also made.
It is shown that quantum gravity effects may prevent
(or, at least, delay or soften) the cosmic doomsday catastrophe
associated with the phantom, i.e.
 the otherwise unavoidable finite-time future singularity (Big Rip).
A novel dark energy model (higher-derivative scalar-tensor theory)
is introduced and it is shown to admit an effective phantom/quintessence
description with a transient acceleration phase. In this case, gravity
favors that an initially insignificant portion of dark energy becomes
dominant over the standard matter/radiation components in the evolution process.

\vfill

\end{abstract}

\noindent
PACS numbers: 98.80.-k,04.50.+h,11.10.Kk,11.10.Wx

\end{titlepage}

\section{Introduction}

Recent astrophysical data, ranging from high redshift surveys of
supernovae to WMAP observations, indicate that about 70 percent of
the total energy of our universe is to be attributed to a weird
cosmic fluid with large and negative pressure, the dark energy
(see \cite{p,expo1} for a recent review) and that the universe is
currently in an accelerating phase. It also turns out that the
dark energy equation of state parameter $w$ is close to $-1$. So
far, the simplest possibility proposed for such kind of dark
energy is the use of a scalar field (or a scalar-tensor theory).
However, scalar-tensor theories are not free from problems,
especially when they are considered directly as dark energy
candidates.

Much attention has been drawn by scalar fields in studies of the
early time universe. A variety of scalar potentials have been
considered and a number of accelerating (inflationary) cosmologies
have been advocated. For instance, the interesting quintessence
model \cite{quint} with $w$ slightly bigger than $-1$ is quite
popular for the explanation of early (and late) time acceleration,
especially in the case of exponential potentials \cite{expo}.
Moreover, exponential scalar potentials often appear naturally
after compactification in string/M-theory. Needless to say, such
description is model-dependent and is still quite far from the
final goal: the formulation of a plausible and consistent dark
energy theory.

Another line of research is related with the case where the dark
energy equation of state parameter is less than $-1$, since this
possibility is not excluded by astrophysical data. The typical
example of a dark energy of this kind is provided by a scalar
field with negative kinetic energy, dubbed phantom (see
\cite{phantom, phantom1} and references therein). At first sight,
such models may look rather strange and they lead to a number of
unpleasant consequences, as a finite-time future singularity (the
Big Rip) \cite{alexei, brett,CKM}. Nevertheless, the possibility
of negative energies seems to be acceptable in classical
scalar-tensor theories. Actually, many of them do contain
phantoms, as the ones coming from string/M-theory
compactification, or higher-derivative supergravities, or
modifications of Einstein gravity itself. In fact, the issue is
somehow delicate, since what looks like a phantom in one reference
frame may radically change its nature in another frame (e.g. after
a conformal transformation). In this sense, even in the absence of
fundamental, physical meaning the phantom can be still useful as a
convenient mathematical tool in order to study cosmological models
in standard scalar-tensor theories, because a phantom-related
frame may lead to a simpler formulation of the problem. Finally,
there are examples where an effective phantom/quintessence
description of the late-time universe naturally appears, even if
the starting theory does not explicitly exhibit the
phantom/quintessence structure.

In the present work we study different cases of a late-time
spatially-flat FRW cosmology in the (phantom) scalar-tensor
theory, mainly with exponential potentials. Such scalar is
considered as a dark energy and the possibility of deriving the
current speed-up is shown also in the presence of matter. Exact
FRW cosmologies are constructed for the (phantom) scalar-tensor
theory with an exponential potential, a model that can be
important for understanding attractors and the stability
properties. The possibility of avoiding the unwanted Big Rip by
simply taking into account quantum gravity effects, which may
become dominant near future singularity, is demonstrated. Finally,
the present dark energy dominance and acceleration, within the
effective phantom/quintessence description, is discussed in the
model where a new form of higher-derivative, gravity-matter
coupling is introduced.

The organization of the paper is the following. In the next
section spatially-flat FRW cosmological solutions are discussed
for scalar-tensor gravity with scalar matter. Explicit examples of
accelerating (and decelerating) scale factors are presented for
exponential potentials when the theory contains one or two scalar
fields. In Sect. 3 the general solution for a spatially-flat FRW
cosmology which includes eternal or transient acceleration is
found in the (phantom) scalar-tensor theory with exponential
potential. This is based on the use and extension of a method
recently developed by J. Russo. The comparison with particular
solutions of the previous section is done. Sect. 4 is devoted to
the study of the influence of quantum gravity effects on the Big
Rip singularity in phantom cosmology. It is shown that taking them
into account properly may change the future of the universe, from
that with a finite-time singularity to an ordinary deSitter space.
In Sect. 5 a higher derivative matter-gravity coupling is
suggested as a new sort of dark energy model. It admits an
effectively phantom/quintessence description and does explain the
current dark energy dominance over standard matter by gravity
assistance. Stability analysis of the model demonstrates that the
acceleration phase is actually transient. A summary and outlook
are given in the discussion. In the Appendix we outline how the
(phantom) scalar-tensor theory may originate, via
compactification, from a higher-dimensional (super)gravity.

\section{Examples of the accelerating universe in (phantom) scalar-tensor
theories.
\label{S2}}

We start from the action of multi-scalar-tensor theory. Several
illustrative examples of the (accelerating) FRW cosmology will be
presented here as simple dark energy models (see \cite{p,expo1}
for a recent review). A scalar field, $\phi$, which may be later
regarded as a phantom, couples with gravity. As is typical in
these models, a second scalar field, $\chi$, is considered. The
string-inspired Lagrangian in the $d$-dimensional spacetime is:
\bea \label{P1} S&=&{1 \over \kappa^2}\int d^dx
\sqrt{-g}\e^{\alpha\phi}\left(R - {\gamma \over 2}
\partial_\mu \phi \partial^\mu \phi - V(\phi)\right) \nn
&& + \int d^dx \sqrt{-g}\left( -{1 \over 2}\partial_\mu \chi
\partial^\mu \chi - U(\chi)\right) \ .
\eea Here $\alpha$ and $\gamma$ are constant parameters and
$V(\phi)$ $\left(U(\chi)\right)$ is the potential for $\phi$
$\left(\chi\right)$. If the constant parameter $\gamma$ is
negative, $\phi$ has a negative kinetic energy and can be regarded
as a phantom \cite{phantom, phantom1}. We should note, however,
that $\gamma$ need not be negative in order to obtain the
accelerated universe as we will later see. As the matter scalar
$\chi$ does not couple with $\phi$ directly, the equivalence
principle is not violated, although the effective gravitational
coupling depends on $\phi$ as $\kappa\e^{-{\alpha \phi \over 2}}$.
One may go to the Einstein frame by the scale transformation \be
\label{P2} g_{\mu\nu}=\e^{-{2\alpha \over d-2}\phi}g_{E\,\mu\nu}\
. \ee In the following,  the quantities in the Einstein frame are
denoted by the index $E$. After the transformation (\ref{P2}), the
action (\ref{P1}) has the following form: \bea \label{P3} S&=&{1
\over \kappa^2}\int d^dx \sqrt{-g_E}\left(R_E -
\left({(d-1)\alpha^2 \over d-2} +{\gamma \over
2}\right)g_E^{\mu\nu}
\partial_\mu \phi \partial_\nu \phi - \e^{-{2\alpha \over d-2}\phi}V(\phi)\right) \nn
&& + \int d^dx \sqrt{-g_E}\left( -{\e^{-\alpha \phi} \over 2} g_E^{\mu\nu}\partial_\mu \chi
\partial_\nu \chi - \e^{-{d\alpha \over d-2}\phi}U(\chi)\right) \ .
\eea In the Einstein frame, there appears a term coupling the
matter $\chi$ with $\phi$. Even if $\gamma$ is negative, when \be
\label{P4} {(d-1)\alpha^2 \over d-2}+{\gamma \over 2}>0\ , \ee the
kinetic energy of $\phi$ becomes positive, as for a usual scalar
field. Hence, the remarkable observation follows that, what is a
phantom in one frame may not be a phantom in a different frame,
especially if the coupling is taken into account.

As first step, one considers the $d=4$ case and  assumes $\chi=0$.
We now define $\varphi$ and $\tilde V(\varphi)$ as \be \label{P5}
\varphi=\phi\sqrt{\alpha^2 + {\gamma \over 3}}\ ,\quad \tilde
V(\varphi) = \e^{-\alpha\phi}V(\phi)\ , \ee and  assume that the
metric has the FRW form in flat space: \be \label{P5b} ds_E^2 = -
dt_E^2 + a_E\left(t_E\right)^2\sum_{i=1,2,3}\left(dx^i\right)^2\ .
\ee Here $t_E$ is the time coordinate in the Einstein frame. When
$\varphi$ only depends on the time coordinate,
 the  FRW equation and $\varphi$ equation follow:
\bea
\label{P6}
3H_E^2 &=& {3 \over 4}\left({d\varphi \over dt_E}\right)^2 + {1 \over 2}
\tilde V(\varphi)\ , \\
\label{P7} 0 &=& 3\left({d^2 \varphi \over dt_E^2} + 3H_E{d\varphi
\over dt_E}\right) + {\tilde V}'(\varphi)\ . \eea Here the Hubble
parameter (in the Einstein frame) is defined by $H_E\equiv {1
\over a_E}{da_E \over dt_E}$. Note that models of such sort may
have a double interpretation: as multi-scalar-tensor theories or
as matter-scalar-tensor theories. In other words, some scalars may
be considered as matter or as part of a gravitational theory. It
has been suggested that such models may describe the inflationary
early universe as quintessence \cite{quint}.

Special attention in cosmology has been paid to exponential
potentials \cite{expo}, which often follow from string/M-theory
compactification. If $\tilde V(\varphi)$ behaves as an exponential
function of $\varphi$ \be \label{P7b} \tilde V(\varphi)\sim V_0
\e^{-2{\varphi \over \varphi_0}} \ee ($V_0$ and $\varphi_0$ are
constants) during some period, as in the present universe, the
solution of (\ref{P6}) and (\ref{P7}) exists:\footnote{The action
(\ref{P1}) with the potential  (\ref{P7b}) belongs to the class
discussed in \cite{Stra}. In fact if we redefine the field $\phi$
by
\[
\Phi\equiv {2\sqrt{|\gamma|} \over \alpha}\e^{{\alpha \over 2}\phi}\ ,
\]
the action (\ref{P1}) can be rewritten as
\begin{eqnarray*}
S&=&{1 \over \kappa^2}\int d^dx \sqrt{-g}\left(F(\Phi)R \mp {1 \over 2}
\partial_\mu \Phi \partial^\mu \Phi - U(\Phi)\right) \nn
&& + \int d^dx \sqrt{-g}\left( -{1 \over 2}\partial_\mu \chi
\partial^\mu \chi - U(\chi)\right) \ .
\end{eqnarray*}
Here
\[
F(\Phi)={\alpha^2 \over 4|\gamma|}\Phi^2\ ,\quad
U(\Phi)=V_0\left({\alpha \Phi \over 2\sqrt{|\gamma|}}\right)^{4\left(1 - \sqrt{1+{\gamma
\over 3\alpha^2}}\right)}\ .
\]
Then, one can use the same arguments as in \cite{Stra} in order to
fit the parameters so that they satisfy the present cosmological
data. The non-minimal scalar-gravitational coupling term which is
required by renormalizability of the quantum field theory in
curved spacetime \cite{BO, BO1} may have very interesting effects
on the phantom cosmology (see \cite{valerio} for a recent
discussion.)} \be \label{P8} a_E=a_{E0}\left({t_E \over
t_{E0}}\right)^{{3 \over 4}\varphi_0^2}\ ,\quad \varphi =
\varphi_0 \ln {t_E \over t_{E0}}\ . \ee Here \be \label{P8b}
t_{E0}\equiv \varphi_0 \sqrt{{1 \over V_0} \left({27 \over
8}\varphi_0^2 - {3 \over 2}\right)}\ . \ee In the original
(Jordan) frame  (\ref{P1}), the time coordinate $t$ and scale
factor $a(t)$ in the FRW form of the metric \be \label{P9} ds^2 =
- dt^2 + a\left(t\right)^2\sum_{i=1,2,3}\left(dx^i\right)^2\ , \ee
are related with the corresponding quantities in the Einstein
frame (\ref{P5b}) by \bea \label{P10} && dt=\e^{-{\alpha \over
2}\phi}dt_E ={t_{E0}^{\beta \varphi_0 \over 2} \over 1 -
{\beta\varphi_0 \over 2}}
d\left( t_E^{1- {\beta \varphi_0 \over 2}}\right) \\
\label{P11}
&& a=\e^{- {\alpha\phi \over 2}}a_E
=a_{E0}\left({t_E \over t_{E0}}\right)^{{3 \over 4}\varphi_0^2 - \beta\varphi_0}
=a_{E0}\left({t \over t_0}\right)^{{3 \over 4}\varphi_0^2 - \beta\varphi_0
\over 1 - {\beta \varphi_0 \over 2}}\\
\label{P11c} && \phi={\beta \varphi_0 \over \alpha \left(1 -
{\beta \varphi_0 \over 2}\right)} \ln {t \over t_0}\ . \eea Here
\be \label{P11b} \beta\equiv {\alpha \over \sqrt{\alpha^2 +
{\gamma \over 3}}}\ ,\quad t_0\equiv {t_{E0} \over 1 - {\beta
\varphi_0 \over 2}}\ . \ee Then, cosmic acceleration ($\dot a>0$
and $\ddot a>0$) occurs when \be \label{P10b} {{3 \over
4}\varphi_0^2 - {\varphi_0 \alpha \over 2\sqrt{\alpha^2 + {\gamma
\over 3}}} \over 1-{\varphi_0 \alpha \over 2\sqrt{\alpha^2 +
{\gamma \over 3}}}}>1\ . \ee By properly choosing the parameters
$\alpha$, $\varphi_0$, and $\gamma$, the present cosmic
acceleration can be realized. For the matter with $p=w\rho$, where
$p$ is the pressure and $\rho$ is the energy density, \be
\label{P12} a\propto t^{2 \over 3(w+1)} \ . \ee Then, effectively
\be \label{P13} w = -1 + {{2 \over 3}\left(1-{\varphi_0 \alpha
\over 2\sqrt{\alpha^2 + {\gamma \over 3}}} \right) \over {3 \over
4}\varphi_0^2 - {\varphi_0 \alpha \over 2\sqrt{\alpha^2 + {\gamma
\over 3}}}} =-{\varphi_0\left(2\beta - 9\varphi_0\right) + 8 \over
3\left(2\beta - 3\varphi_0\right) \varphi_0}\ . \ee As $w$
diverges when the denominator in the second term vanishes, $w$ can
take any value by properly choosing $\alpha$, $\varphi_0$, and
$\gamma$. For example, if \be \label{P14} {\varphi_0 \alpha \over
\sqrt{\alpha^2 + {\gamma \over 3}}}=2\ , \ee
 $w=-1$. Several interesting cases deserve attention:
\bea
\label{P15}
&& \mbox{when}\ \varphi_0^2>{4 \over 3} \nn
&& \left\{\begin{array}{lll}
w<-1 &\mbox{if}& {3 \over 2}\varphi_0^2 >
{\varphi_0 \alpha \over \sqrt{\alpha^2 + {\gamma \over 3}}} > 2 \\
w>-1 &\mbox{if}& {\varphi_0 \alpha \over \sqrt{\alpha^2 + {\gamma \over 3}}}
> {3 \over 2}\varphi_0^2
\ \mbox{or}\ {\varphi_0 \alpha \over \sqrt{\alpha^2 + {\gamma \over 3}}} < 2 \ ,
\end{array}\right. \\
\label{P16}
&& \mbox{when}\ \varphi_0^2<{4 \over 3}  \nn
&& \left\{\begin{array}{lll}
w<-1 &\mbox{if}& 2 > {\varphi_0 \alpha \over \sqrt{\alpha^2 + {\gamma \over 3}}}
> {3 \over 2}\varphi_0^2 \\
w>-1 &\mbox{if}& {\varphi_0 \alpha \over \sqrt{\alpha^2 + {\gamma \over 3}}}
< {3 \over 2}\varphi_0^2
\ \mbox{or}\ {\varphi_0 \alpha \over \sqrt{\alpha^2 + {\gamma \over 3}}}> 2 \ .
\end{array}\right.
\eea As is clear from (\ref{P12}), cosmic acceleration
$\left(\ddot a>0\right)$ occurs when \be \label{P17} w<-{1 \over
3}\ . \ee When $w<-1$, the universe is not expanding but
shrinking, although the universe is still accelerating. In the
case $w<-1$, if we change the direction of time as $t\to t_s - t$,
the universe is both accelerating and expanding.

In the especial case when $w=-{1 \over 3}$, we find \be
\label{P18} \varphi_0^2={4 \over 3}\ . \ee Note that even when
$\gamma$ is positive ---e.g. the kinetic energy of $\phi$ is
positive as for usual matter--- the effective $w$ can still be
less than $-1$. For example, with the choice \be \label{Pex1}
\varphi_0=4\ ,\quad {\gamma \over \alpha^2}=3\ , \ee it follows
that $\beta={1 \over \sqrt{2}}$, and  from (\ref{P13}), \be
\label{Pex2} w=- {11\sqrt{2} + 203 \over 213}=-1.025...<-1\ . \ee

Note that the above considerations are appliable both for the
early as well as for the late time universe. Indeed, the simple
solution under discussion (especially, the no phantom case) has
been considered in different situations. Later on, the general
solution for the scalar-tensor theory with an exponential
potential will be discussed. With a known general solution, it is
much easier to understand the type of situation which appears: if
it is a transient (or eternal) acceleration, or an attractor, or
something else.

In the previous example the matter field $\chi$ is zero
$\left(\chi=0\right)$. We now consider the case of $\chi\neq 0$
which is much related with the so-called double quintessence
model\cite{double}. In the Einstein frame (\ref{P3}), the FRW
equation, $\varphi$ equation, and $\chi$ equation have the
following form: \bea \label{P19} 3H_E^2 &=& {3 \over
4}\left({d\varphi \over dt_E}\right)^2 + {1 \over 2} \tilde
V(\varphi) + {\kappa^2 \over 4}\e^{-\beta\varphi}\left({d\chi
\over dt_E}\right)^2
+ {\kappa^2 \over 2}\e^{-2\beta\varphi}U(\chi) \ , \\
\label{P20}
0 &=& 3\left({d^2 \varphi \over dt_E^2} + 3H_E{d\varphi \over dt_E}\right)
+ {\tilde V}'(\varphi) + {\beta \kappa^2 \over 2}
\e^{-\beta\varphi}\left({d\chi \over dt_E}\right)^2 \nn
&& -2\beta \kappa^2 \e^{-2\beta\varphi}U(\chi)\ , \\
\label{P21} 0 &=& {d \over dt_E}\left(\e^{-\beta\varphi}{d\chi
\over dt_E}\right) + 3H_E\e^{-\beta\varphi}{d\chi \over dt_E}+
\e^{-2\beta\varphi}U'(\chi)\ . \eea We again assume the
exponential potentials \be \label{P31} \tilde V(\varphi)=V_0
\e^{-{2 \over \varphi_0}\varphi}\ ,\quad U(\chi)=U_0\chi^{4-{4
\over \beta\varphi_0}}\ . \ee The following Ansatz exists \be
\label{P32} H_E={h_0 \over t_E}\ ,\quad \varphi=\varphi_0 \ln {t_E
\over t_{E0}}\ ,\quad \chi=\chi_0\left({t_E \over
t_{E0}}\right)^{\beta\varphi \over 2}\ . \ee The parameters $h_0$,
$t_{E0}$, and $\chi_0$ should be a solution of the following
algebraic equations, which can be obtained from (\ref{P19}),
(\ref{P20}), (\ref{P21}): \bea \label{P33} 0&=& - 3h_0^2 + {3
\over 4}\varphi_0^2 + {1 \over 2}V_0 t_{E0}^2 + {\kappa^2 \beta^2
\varphi_0^2 \chi_0^2 \over 16} + {\kappa^2 \over 2}U_0 t_{E0}^2
\chi_0^{4 - {4 \over \beta\varphi_0}}\ ,\nn 0&=& 3\left(-\varphi_0
+ 3h_0\varphi_0\right) - {2V_0 t_{E0}^2 \over \varphi_0} +
{\kappa^2 \beta^3 \varphi_0^2 \chi_0^2 \over 8}
 - 2\kappa^2 \beta U_0 t_{E0}^2 \chi_0^{4 - {4 \over \beta \varphi_0}} \ ,\nn
0&=& -{\beta \varphi_0 \over 2}\left({\beta\varphi_0 \over 2} +
1\right) + {3h_0\beta \varphi_0 \over 2} + 4U_0t_{E0}^2\left(1 -
{1 \over \beta\varphi_0}\right) \chi_0^{2 - {4 \over \beta
\varphi_0}} \ . \eea For instance, for the special example \be
\label{P34} \varphi_0=\sqrt{10 \over 27}\ ,\quad \beta=2\sqrt{27
\over 10}\ ,\quad V_0=0\ , \ee the explicit solution follows: \be
\label{P35} h_0={5 \over 12}\ ,\quad \chi_0={\sqrt{5} \over
3\kappa}\ ,\quad t_{E0}=\sqrt{3 \over 8U_0}\ . \ee Using
(\ref{P11b}) and (\ref{P34}), one arrives at \be \label{P34b}
\gamma = -{49 \over 18}\alpha^2\ . \ee Then $\phi$ is surely a
phantom with negative kinetic energy in the physical Jordan frame
(\ref{P1}). Since  $U(\chi)=U_0\chi^2$, $U(\chi)$ corresponds to a
mass term and the mass $m_\chi$ is given by \be \label{P36}
m_\chi^2 = 2 U_0\ . \ee Also \be \label{P37} t_{E0}={\sqrt{3}
\over 2m_\chi}\ . \ee Since $dt=\pm \e^{-{\alpha \phi \over
2}}dt_E=\pm \e^{-{\beta \varphi \over 2}}dt_E = \pm {t_{E0} \over
t_E}dt_E$, we find \be \label{P38} {t \over t_{E0}}=\pm \ln {t_E
\over t_{E0}}\ . \ee Then, in the physical Jordan frame one gets
\bea \label{P39} && a=\e^{-{\alpha \phi \over 2}}a_E =
a_{E0}\left({t_E \over t_{E0}}\right)^{-{7 \over 12}}
= a_{E0}\e^{\mp {7t \over 12 t_{E0}}}\ ,\\
&& \phi ={\beta \over \alpha}\varphi = {2 \over \alpha}\ln {t_E \over t_{E0}}
=\pm {2 \over \alpha}{t \over t_{E0}}\ ,\\
&& \chi = \chi_0{t_E \over t_{E0}}=\chi_0 \e^{\pm {t \over
t_{E0}}}\ . \eea Since the scale factor $a$ behaves as an
exponential function of $t$, the Hubble parameter $H$ is a
constant \be \label{P40} H=\mp {7 \over 12t_{E0}}=\mp {7m_\chi
\over 6\sqrt{3}}\ , \ee which is linear in the mass $m_\chi$ of
$\chi$. Eq.~ (\ref{P40}) also tells that the effective $w$ is $-1$
(similar to the cosmological constant).

As a different, more complicated example one can consider the case
when $\alpha=0$, and therefore $\beta=0$, but the potential
depends on both $\phi$ and $\chi$: \be \label{PB1} S=\int d^4 x
\sqrt{-g}\left\{{1 \over \kappa^2}\left(R - {3 \over 2}
\partial_\mu \varphi \partial^\mu \varphi\right)  -{1 \over 2}\partial_\mu \chi
\partial^\mu \chi - W(\varphi,\chi)\right\} \ .
\ee As $\alpha=0$, the Einstein frame can be regarded as a truly
physical one. Then the equations corresponding to (\ref{P19}),
(\ref{P20}),
 and (\ref{P21}) are:
\bea
\label{PB2}
3H^2 &=& {3 \over 4}\left({d\varphi \over dt}\right)^2
+ {\kappa^2 \over 4}\left({d\chi \over dt}\right)^2
+ {\kappa^2 \over 2}W(\varphi, \chi) \ , \\
\label{PB3-0}
0 &=& 3\left({d^2 \varphi \over dt^2} + 3H{d\varphi \over dt}\right)
+ \kappa^2 W_{,\varphi}(\varphi,\chi)\ , \\
\label{PB4-0} 0 &=& {d^2\chi \over dt^2}+ 3H{d\chi \over dt}+
W_{,\chi}(\varphi, \chi)\ . \eea Here   the derivative of
$W(\varphi,\chi)$ with respect to $\varphi$ ($\chi$) is expressed
as $W_{,\varphi}\left(\varphi,\chi\right)$
$\left(W_{,\chi}\left(\varphi,\chi\right)\right)$. The case of the
exponential potential  $W(\varphi,\chi)$ may be of interest \be
\label{PB3} W(\varphi,\chi)=W_0 \e^{{\eta \over \chi_0}\chi - {2 +
\eta \over \varphi_0}\varphi}\ , \ee where $W_0$, $\eta$,
$\chi_0$, and $\varphi_0$ are constant parameters. Assuming \be
\label{PB4} H={h_0 \over t}\ ,\quad \varphi=\varphi_0 \ln {t \over
t_0}\ ,\quad \chi = \chi_0 \ln {t \over t_0}\ , \ee with constants
$h_0$ and $t_0$, Eqs.~(\ref{PB2}), (\ref{PB3-0}), and
(\ref{PB4-0}) reduce to the following algebraic equations: \bea
\label{PB5} 0&=& - 3h_0^2 + {3 \varphi_0^2 \over 4} + {\kappa^2
\chi_0^2 \over 4} + {\kappa^2 W_0 t_0^2 \over 2} \nn 0&=&
3\varphi_0^2 \left(-1 + 3h_0\right) - \left(2+\eta\right)\kappa^2
W_0 t_0^2\ ,\nn 0&=& 3\chi_0^2 \left(-1 + 3h_0\right) +
\eta\kappa^2 W_0 t_0^2\ . \eea Eqs.~ (\ref{PB5}) give \bea
\label{PB6}
&& \kappa^2 \chi_0^2 = - {3\eta \over 2+\eta}\varphi_0^2\ ,\\
\label{PB7}
&& h_0 = {3\varphi_0^2 \over 2(2+\eta)}\ ,\\
\label{PB8-0} && \kappa^2 W_0 t_0^2 = {3\varphi_0^2\left(3h_0 -
1\right) \over 2+ \eta} \ . \eea Eq.~(\ref{PB6}) shows that \be
\label{PB8} -2\leq \eta\leq 0\ . \ee The effective $w$ is given by
\be \label{PB9} w=-1 + {2 \over 3h_0} = -1 + {4(2+\eta) \over
9\varphi_0^2}\ . \ee As discussed in (\ref{PB8}), as $w>-1$  we do
not have a phantom. If the second term in (\ref{PB9}) is small,
one may obtain quintessence.

Other types of matter can be easily considered too. For instance,
matter may be dust.
 The energy density
$\rho_{\rm dust}$ in the Jordan frame  (\ref{P1}) behaves as
$\rho_{\rm dust}\propto a^{-3}$. Then \be \label{PD1} \rho_{\rm
dust}=\rho_0 a^{-3}=\rho_0 \e^{{3\alpha \over 2}\phi}a_E^{-3}
=\rho_0 \e^{{3\beta \over 2}\varphi}a_E^{-3}\ ,\quad p_{\rm
dust}=0\ . \ee It is well known that dust has no pressure. It
could correspond to the baryon and/or cold dark matter components.
Instead of (\ref{P19}) and (\ref{P20}), we obtain the following
equations of motion in the Einstein frame: \bea \label{PD2} 3H_E^2
&=& {3 \over 4}\left({d\varphi \over dt_E}\right)^2 + {1 \over 2}
\tilde V(\varphi) + {\kappa^2 \over 2}\e^{-{\beta \over 2}\varphi}
\rho_0 a_E^{-3} \ , \\
\label{PD3} 0 &=& 3\left({d^2 \varphi \over dt_E^2} +
3H_E{d\varphi \over dt_E}\right) + {\tilde V}'(\varphi) -{\beta
\kappa^2 \over 2}\e^{-{\beta \over 2}\varphi} \rho_0 a_E^{-3}\ .
\eea With the form of $\tilde V(\varphi)$ as in (\ref{P31}), the
solution occurs \be \label{RD3} \varphi=\varphi_0 \ln {t_E \over
t_{E0}}\ ,\quad a_E=a_0\left({t_E \over t_{E0}}\right)^{{2 \over
3}- {\beta\varphi_0 \over 6}}\ . \ee Here \bea \label{RD4} t_0^2
&=& {\beta^2 \varphi_0^2 + 9 \varphi_0^2 - 9\beta\varphi_0 \over
6V_0} \ ,\nn a_0^3 &=& {\rho_0 \kappa^2 \e^{-{\beta\varphi_0 \over
2}} \left(\beta^2 \varphi_0^2 + 9 \varphi_0^2 -
9\beta\varphi_0\right) \over 2V_0 \left(8 - 2\beta \varphi_0 - 9
\varphi_0^2 \right)}\ . \eea The time-coordinate $t$ in the Jordan
frame is given by \be \label{RD5} {t \over t_0}=\left({t_E \over
t_{E0}}\right)^{1 - {\beta\varphi_0 \over 2}}\ , \quad t_0\equiv
\left|{2t_{E0} \over \beta \varphi_0}\right|\ , \ee and the scale
factor can be obtained: \be \label{RD6} a=a_0\left({t \over
t_0}\right)^{{2 \over 3}{2\left(1 -\beta \varphi_0\right) \over 2
- \beta\varphi_0}}\ . \ee The effective $w$ is found to be \be
\label{RD7} w=-1 + {2 - \beta\varphi_0 \over 2\left(1-
\beta\varphi_0\right)}\ . \ee Hence, if \be \label{RD8}
1<\beta\varphi_0 <2\ , \ee
 $w<-1$. If we assume $V_0>0$ (and $t_0^2>0$ and $\rho_0 a_0^{-3}>0$),
Eqs.~ (\ref{RD4}) require \be \label{RD9} \beta^2 \varphi_0^2 + 9
\varphi_0^2 - 9\beta\varphi_0 >0\ ,\quad 8 - 2\beta \varphi_0 - 9
\varphi_0^2>0\ , \ee which give $\beta^2\varphi_0^2 - 11\beta
\varphi_0 + 8>$, that is, \be \label{RD10} \beta\varphi_0<{11 -
\sqrt{89} \over 2}\ \mbox{or}\ \beta\varphi_0>{11 + \sqrt{89}
\over 2} \ . \ee Numerically, $\beta\varphi_0<0.7830...$ or
$\beta\varphi_0>14.9339...$, which contradict the results in
(\ref{RD8}), and there is no accelerating universe. With the
assumption $V_0<0$, the accelerating universe takes over. For
instance, with the choice $\beta\varphi_0={3 \over 2}$ and
$\varphi_0^2={1 \over 9}$ ($\beta = \pm {9 \over 2}$ and
$\varphi_0=\pm {1 \over 3}$), we obtain $t_0^2 = -{41 \over 24
V_0}$ and $a_0^3 = - {\rho_0\kappa^2 \e^{-{\beta\varphi_0 \over
2}} \over 2V_0}$.

Going back to the case of two scalars, by variation over
$g_{\mu\nu}$ the Einstein equation follows: \bea \label{P27} && {1
\over \kappa^2}\left( R_{\mu\nu} - {1 \over 2}g_{\mu\nu} R \right)
= {1 \over 2}\left(T^\phi_{\mu\nu} + T^\chi_{\mu\nu}\right) \ ,\nn
&& T^\phi_{\mu\nu}\equiv {1 \over \kappa^2}\left( - {\gamma \over
2}\partial_\rho \phi
\partial^\rho \phi g_{\mu\nu} + \gamma \partial_\mu \phi \partial_{\nu} \phi
 - V(\phi)g_{\mu\nu} \right. \nn
&& \quad \left. + 2\e^{-\alpha\phi} \nabla_\mu \nabla_\nu \left(\e^{\alpha\phi}\right)
 - 2g_{\mu\nu} \e^{-\alpha\phi} \nabla^2 \left(\e^{\alpha\phi}\right)\right) \ ,\nn
&& T^\chi_{\mu\nu} = \e^{-\alpha\phi}\left( - {1 \over 2}\partial_\rho \chi
\partial^\rho \chi g_{\mu\nu} + \partial_\mu \chi \partial_{\nu} \chi
 - U(\chi) g_{\mu\nu}\right) \ .
\eea
 $T^\phi_{\mu\nu}$ and $T^\chi_{\mu\nu}$ may be regarded as the effective
energy-momentum tensor of $\phi$ and $\chi$,
respectively\footnote{The usual energy-momentum tensors ${\cal
T}^{\phi,\chi}_{\mu\nu}$ given
 by the variation over
$g_{\mu\nu}$ are related with $T^{\phi,\chi}_{\mu\nu}$ by ${\cal
T}^{\phi,\chi}_{\mu\nu}=\e^{\alpha\phi} T^{\phi,\chi}_{\mu\nu}$.}.
In particular, in the FRW metric (\ref{P9}), we find \bea
\label{P28} T^\phi_{tt} &=& \rho^\phi \nn &=& {1 \over
\kappa^2}\left\{ {\gamma \over 2}{\dot \phi}^2 + V(\phi)
 - 6\alpha H \dot \phi\right\} \ ,\nn
T^\chi_{tt} &=& \rho^\chi \nn
&=& \e^{-\alpha\phi}\left\{ {1 \over 2}{\dot \chi}^2 + U(\chi)\right\} \ ,\nn
T^\phi_{ij}&=& p^\phi a^2 \delta_{ij} \nn
&=& {1 \over \kappa^2}\left\{ {\gamma \over 2}{\dot \phi}^2 - V(\phi)
+ 2\alpha \ddot \phi + 2\alpha^2 {\dot \phi}^2
+ 4\alpha H \dot\phi \right\}a^2 \delta_{ij} \ ,\nn
T^\chi_{ij}&=& p^\chi a^2 \delta_{ij} \nn
&=& \e^{-\alpha\phi}\left\{ {1 \over 2}{\dot \chi}^2 - U(\chi)\right\}a^2 \delta_{ij}\ .
\eea
The effective $w^\phi$ and $w^\chi$ can be defined as follows:
\be
\label{P29}
w^\phi = {p^\phi \over \rho^\phi}\ ,\quad
w^\chi = {p^\chi \over \rho^\chi}\ .
\ee

For the first example in (\ref{P10}), (\ref{P11}), and
(\ref{P11c}), it follows that \bea \label{P30} \rho^\phi &=& {3
\left(2\beta - 3\varphi_0\right)^2 \varphi_0^2 \over 8 \kappa^2
t^2 \left(1 - {\beta \varphi_0 \over 2}\right)^2} \ , \nn p^\phi
&=& { \left(2\beta - 3\varphi_0\right)\left\{ \varphi_0^2
\left(2\beta - 9\varphi_0\right) + 8\varphi_0\right\} \over 8
\kappa^2 t^2 \left(1 - {\beta \varphi_0 \over 2}\right)^2} \ ,
\eea which agrees with $w={p \over \rho}={p_\phi \over \rho_\phi}$
in (\ref{P13}). Note that the consideration of various entropies
for dark energy models can be done and
 then interesting holographic relations among them occur (see
\cite{brevik} for a recent discussion).

For the second example in (\ref{P39}), one gets \bea \label{P41}
&& \rho^\phi= {14 \over 9\kappa^2 t_{E0}^2}\ ,\quad p^\phi= -{19
\over 9\kappa^2 t_{E0}^2}\ , \nn && \rho^\chi= {35 \over
72\kappa^2 t_{E0}^2}\ ,\quad p^\phi= {5 \over 72\kappa^2
t_{E0}^2}\ . \eea Hence, \be \label{P42} w^\phi=-{19 \over 14}\
,\quad w^\chi={1 \over 7}\ . \ee However, since \be \label{P43}
\rho^\phi + \rho^\chi = -\left(p^\phi + p^\chi\right)={147 \over
72 \kappa^2 t_{E0}^2}\ , \ee
 it turns out that $w=-1$, which is consistent with (\ref{P40}). We should also note that, in the
Einstein frame (\ref{P3}), $\phi$ has a positive kinetic energy.

For the case of (\ref{PB4}) with (\ref{PB6}), (\ref{PB7}), and (\ref{PB8-0}) we find
\bea
\label{P44}
&& \rho = {3 \over 2\kappa^2}{\dot \varphi}^2 + {1 \over 2}{\dot \chi}^2 + W(\phi,\chi)
= {18\varphi_0^2 h_0 \over 2t^2\kappa^2(2+\eta)} \ ,\nn
&& p = {3 \over 2\kappa^2}{\dot \varphi}^2 + {1 \over 2}{\dot \chi}^2 - W(\phi,\chi)
= {6\varphi_0^2 \left(2 - 3h_0\right) \over 2t^2\kappa^2(2+\eta)} \ ,
\eea
which reproduces (\ref{PB9}).
Having these various examples of the (accelerated) evolution of the
current universe one
can compare it with recent astrophysical data in the way discussed
recently in \cite{expo1,NP}. Of course, above illustrative examples of
current speed-up may correspond to transient acceleration.
In other words, stability of the solutions pretending to be realistic ones
should be investigated in detail.

\section{Exact FRW cosmology for the (phantom) scalar-tensor theory
with an exponential potential. }

In the present section the exact FRW cosmology in the (phantom)
scalar-tensor theory with an exponential potential will be
discussed. First, the method which is appropriate to obtain the
exact FRW solutions will be reviewed \cite{russo} (for
introduction of similar variables in quantum cosmology,
see \cite{james}).  Subsequently,
 the formulation is extended to the phantom case with negative kinetic
term.

The action of a scalar field $\phi$ coupled with gravity is: \be
\label{R1} S={1 \over \kappa^2}\int d^4 x \sqrt{-g} \left( R -
{\gamma \over 2}
\partial_\mu \phi \partial^\mu \phi - V(\phi) \right)\ .
\ee
This action can be regarded as the action with $\alpha=0$ in (\ref{P1}) or
that obtained by replacing ${(d-1)\alpha^2 \over d-2}+{\gamma \over 2}$
and $\e^{-{2\alpha \over d-2}\phi}V(\phi)$ in (\ref{P3}) with
${\gamma \over 2}$ and $V(\phi)$, respectively.
For the standard scalar $\gamma>0$ and one can normalize $\phi$ to be
$\gamma=1$ but for the phantom field with negative kinetic term, we have
$\gamma<0$.

For the FRW metric \be \label{R2} ds^2 = -dt^2 + a(t)^2
\sum_{i=1,2,3}\left(dx^i\right)^2\ , \ee the action (\ref{R1}) can
be rewritten as \be \label{R3} S={1 \over \kappa^2}\int d^3 x dt
\left\{ -6a{\dot a}^2 + a^3 \left({\gamma \over 2} {\dot \phi}^2 -
V(\phi)\right)\right\}\ . \ee The potential $V(\phi)$ is chosen to
be \be \label{R4} V(\phi)=V_0 \e^{-{2\phi \over \phi_0}}\ , \ee
with constants $V_0$ and $\phi_0$.

First we review the standard case with $\gamma>0$ following to
\cite{russo}. The field variables $a$ and $\phi$ are written in
terms of  new fields $v$ and $u$ as \be \label{R5} a=\e^{v+u \over
3}\ ,\quad \phi ={2(v-u) \over \sqrt{3\gamma}}\ , \ee and a new
time variable $\tau$ is defined by \be \label{R6} d\tau = dt
\sqrt{3V_0 \over 8}\e^{-{2(v-u) \over \phi_0\sqrt{3\gamma}}}\ .
\ee Then, the action (\ref{R3}) acquires the following form: \be
\label{R7} S=-{1 \over \kappa^2}\sqrt{8V_0 \over 3}\int d^3 x
d\tau \left[ {dv \over d\tau}{du \over d\tau} + 1 \right]\e^{v+u -
{\bar \alpha} (v-u)}\ ,\quad {\bar \alpha}\equiv {4 \over
\phi_0\sqrt{3\gamma}}\ . \ee Varying over $v$ and $u$,  the
equations of motion follow: \bea \label{R8b} 0&=&{d^2 u \over
d\tau^2} + (1+{\bar \alpha})\left({du \over d\tau}\right)^2 -
\left(1-{\bar \alpha}\right)\ ,
\\
0&=&{d^2 v \over d\tau^2} + (1-{\bar \alpha})\left({du \over
d\tau}\right)^2 - \left(1+{\bar \alpha}\right)\ . \eea Since the
Hamiltonian $H$ conjugate to $\tau$ is given by \be \label{R9}
H=-{1 \over \kappa^2}\sqrt{8V_0 \over 3}\int d^3 x \left[ {dv
\over d\tau}{du \over d\tau} - 1 \right]\e^{v+u - {\bar \alpha}
(v-u)}\ , \ee the Hamiltonian constraint $H=0$ yields \be
\label{R10b} {dv \over d\tau}{du \over d\tau} =1\ . \ee In terms
of the new variables ${\cal V}$ and ${\cal U}$, which are given by
\be \label{R11} {\cal V}\equiv \e^{(1 - {\bar \alpha})v}\ ,\quad
{\cal U}\equiv \e^{(1+{\bar \alpha})u}\ , \ee the equations of
motion are \be \label{R12b} {d^2 {\cal U} \over
d\tau^2}=\left(1-{\bar \alpha}^2\right){\cal U}\ ,\quad {d^2 {\cal
V} \over d\tau^2}=\left(1-{\bar \alpha}^2\right){\cal V}\ . \ee

When $|{\bar \alpha}|<1$, the solution of (\ref{R12b}) is given by
\be \label{R13} {\cal U}=u_+ \e^{\tau\sqrt{1 - {\bar \alpha}^2}} +
u_- \e^{-\tau\sqrt{1 - {\bar \alpha}^2}}\ ,\quad {\cal V}=v_+
\e^{\tau\sqrt{1 - {\bar \alpha}^2}} + v_- \e^{-\tau\sqrt{1 - {\bar
\alpha}^2}}\ , \ee with constants of the integration $u_\pm$ and
$v_\pm$. The Hamiltonian constraint (\ref{R10b}) restricts the
constants as \be \label{R14} u_+ v_-=- u_- v_+\ . \ee Then the
spacetime metric has the following form\cite{russo}: \bea
\label{Ra1} ds^2 &=& -{8 \over 3V_0} \left(v_+ \e^{\tau\sqrt{1 -
{\bar \alpha}^2}} + v_- \e^{-\tau\sqrt{1 - {\bar \alpha}^2}}
\right)^{{\bar \alpha} \over 1 - {\bar \alpha}} \nn && \times
\left(u_+ \e^{\tau\sqrt{1 - {\bar \alpha}^2}} + u_-
\e^{-\tau\sqrt{1 - {\bar \alpha}^2}} \right)^{-{{\bar \alpha}
\over 1+{\bar \alpha}}} d\tau^2 \nn && + \left(v_+ \e^{\tau\sqrt{1
- {\bar \alpha}^2}} + v_- \e^{-\tau\sqrt{1 - {\bar \alpha}^2}}
\right)^{2 \over 3(1 - {\bar \alpha})} \nn && \times \left(u_+
\e^{\tau\sqrt{1 - {\bar \alpha}^2}} + u_- \e^{-\tau\sqrt{1 - {\bar
\alpha}^2}} \right)^{2 \over 3(1+{\bar
\alpha})}\sum_{i=1,2,3}\left(dx^i\right)^2\ . \eea When $\tau\to
+\infty$, Eqs.~ (\ref{R5}) and (\ref{R6}) give \bea \label{Ra2}
a&\to& v_+^{1 \over 3(1-{\bar \alpha})} u_+^{1 \over 3(1+{\bar
\alpha})}\e^{2\tau \over 3\sqrt{1 -{\bar \alpha}^2}}\ , \nn t&\to&
t_{0+} + {\sqrt{1-{\bar \alpha}^2} \over {\bar \alpha}^2}\sqrt{8
\over 3V_0} v_+^{{\bar \alpha} \over 2(1-{\bar
\alpha})}u_+^{-{{\bar \alpha} \over 2(1+{\bar \alpha})}}\e^{{\bar
\alpha}^2 \tau \over \sqrt{1-{\bar \alpha}^2}}\ . \eea Here
$t_{0+}$ is a constant of integration. Hence, $a\propto t^{2 \over
3{\bar \alpha}^2}$ and the universe is accelerating $\left({d^2 a
\over dt^2} > 0\right)$ if \be \label{Ra3} {\bar \alpha}^2< {1
\over 3}\ . \ee Note that if $v_-=u_-=0$, the behavior in
(\ref{Ra2}) is exact even if $\tau$ is not large. The case with
$v_-=u_-=0$ corresponds to the solution in Sect. \ref{S2}. On the
other hand, when $\tau\to -\infty$ \bea \label{Ra4} a&\to& v_-^{1
\over 3(1-{\bar \alpha})} u_-^{1 \over 3(1+{\bar \alpha})}
\e^{-{2\tau \over 3\sqrt{1 -{\bar \alpha}^2}}}\ , \nn t&\to&
t_{0-} - {\sqrt{1-{\bar \alpha}^2} \over {\bar \alpha}^2}\sqrt{8
\over 3V_0} v_-^{{\bar \alpha} \over 2(1-{\bar
\alpha})}u_-^{-{{\bar \alpha} \over 2(1+{\bar \alpha})}}
\e^{-{{\bar \alpha}^2 \tau \over \sqrt{1-{\bar \alpha}^2}}}\ .
\eea Here $t_{0-}$ is again a constant of integration. Thus, we
find $a\propto \left(-t\right)^{2 \over 3{\bar \alpha}^2}$, then
the universe is shrinking but accelerating $\left({d^2 a \over
dt^2} > 0\right)$ if (\ref{Ra3}) is satisfied. To summarize the
$|{\bar \alpha}|<1$ case, if $v_-=u_-=0$ we find an eternal
expanding solution as in Section \ref{S2}.
 In the general case,  the solution is a bouncing universe, where
first the universe shrinks and, after that, it expands.

When$|{\bar \alpha}|>1$, the solution of (\ref{R12b}) can be
written as \bea \label{R15} {\cal U}&=&u_c
\cos\left(\tau\sqrt{{\bar \alpha}^2-1}\right) + u_s
\sin\left(\tau\sqrt{ {\bar \alpha}^2 - 1}\right)\ ,\nn {\cal
V}&=&v_c \cos\left(\tau\sqrt{ {\bar \alpha}^2 - 1}\right) + v_s
\sin\left(\tau\sqrt{ {\bar \alpha}^2 - 1}\right)\ , \eea with
constants $u_c$, $u_s$, $v_c$, and $v_s$,
 which satisfy
\be \label{R15bb} v_c u_c + v_s c_s = 0\ . \ee The solution of
(\ref{R15bb}) can be given by means of three independent
parameters $v_0$, $u_0$, $\theta_0$ as \be \label{Rc1} v_s = v_0
\sin\theta_0\ ,\quad v_c=v_0\cos\theta_0\ ,\quad u_s = u_0
\cos\theta_0\ ,\quad u_c=-u_0\sin\theta_0\ . \ee As a result,
Eq.~(\ref{R15}) simplifies \be \label{Rc2} {\cal
V}=v_0\cos\left(\tau \sqrt{{\bar \alpha}^2 - 1} - \theta_0\right)\
,\quad {\cal U}=u_0\sin\left(\tau \sqrt{{\bar \alpha}^2 - 1} -
\theta_0\right)\ . \ee As $\theta_0$ can be absorbed into the
constant shift of $\tau$, in the following we choose $\theta_0=0$.
The spacetime metric looks as: \bea \label{Rc3} ds^2 &=& - {8
\over 3V_0}v_0^{-{2{\bar \alpha} \over 1-{\bar
\alpha}}}u_0^{2{\bar \alpha} \over 1+ {\bar \alpha}}
\cos^{-{2{\bar \alpha} \over 1-{\bar
\alpha}}}\left(\tau\sqrt{{\bar \alpha}^2 - 1}\right) \sin^{2{\bar
\alpha} \over 1+{\bar \alpha}}\left(\tau\sqrt{{\bar \alpha}^2 -
1}\right) d\tau^2 \nn && + v_0^{2{\bar \alpha} \over 3(1-{\bar
\alpha})}u_0^{2{\bar \alpha} \over 1+ {\bar \alpha}} \cos^{2 \over
3(1-{\bar \alpha})}\left(\tau\sqrt{{\bar \alpha}^2 - 1}\right)
\sin^{2 \over 3(1+{\bar \alpha})}\left(\tau\sqrt{{\bar \alpha}^2 -
1}\right) \ . \eea There are singularities when \be \label{Rc4}
\tau\sqrt{{\bar \alpha}^2 - 1}=n\pi\ ,\quad \mbox{or} \quad
\left(n + {1 \over 2}\right)\pi \ . \ee Here $n$ is an integer. If
we write $\tau$ as $\tau \sqrt{{\bar \alpha}^2 - 1}= n\pi + \delta
\tau$ and assume $\delta \tau$ is small, we find, by neglecting
numerical factors, \be \label{Rc5} t\sim
\left(\delta\tau\right)^{2{\bar \alpha} + 1 \over 1 + {\bar
\alpha}}\ ,\quad a \sim \left(\delta\tau\right)^{1 \over 3(1 +
{\bar \alpha})} \sim t^{1 \over 3(2{\bar \alpha} + 1)}\ . \ee Note
that ${2{\bar \alpha} + 1 \over 1 + {\bar \alpha}}>0$
 as  $|{\bar \alpha}|>1$.
Then $\tau=0$ corresponds to $t=0$. At $t=0$, the size of the
universe diverges (vanishes) when $2{\bar \alpha} + 1<0$ ($2{\bar
\alpha} + 1>0$). On the other hand, if we write $\tau$ as $\tau
\sqrt{{\bar \alpha}^2 - 1}= \left(n + {\pi \over 2} \right)\pi +
\delta \tau$ and assume $\delta \tau$ is small, we find, again
neglecting numerical factors, \be \label{Rc6} t\sim
\left(\delta\tau\right)^{1 - 2{\bar \alpha} \over 1 - {\bar
\alpha}}\ ,\quad a \sim \left(\delta\tau\right)^{1 \over 3(1 -
{\bar \alpha})} \sim t^{1 \over 3(1 - 2{\bar \alpha})}\ . \ee Note
that ${1 - 2{\bar \alpha} + 1 \over 1 - {\bar \alpha}}>0$ once
more, and $\tau=0$ corresponds to $t=0$. Then, at $t=0$ the size
of the universe diverges (vanishes) when $1 - 2{\bar \alpha} <0$
($1 - 2{\bar \alpha} >0$). These results for the standard scalar
with $\gamma>0$ are given in \cite{russo}.

We now extend the above formulation to the case of a phantom with
$\gamma<0$. For this situation, we define a complex field $z$ and
its complex conjugate $z^*$ by \be \label{R15b} a=\e^{z + z^*
\over 3}\ , \quad \phi=-{2i\left(z - z^*\right) \over
\sqrt{-3\gamma}}\ , \ee and define a new time variable $\tau$ as
in (\ref{R6}) by \be \label{R16} d\tau =  \pm dt \sqrt{3V_0 \over
8}\e^{-{2i(z-z^*) \over \phi_0\sqrt{-3\gamma}}}\ . \ee The action
(\ref{R3}) becomes: \be \label{R17} S=\mp {1 \over
\kappa^2}\sqrt{8V_0 \over 3}\int d^3 x d\tau \left[ {dz \over
d\tau}{dz^* \over d\tau} + 1 \right]\e^{z+z^* - i\tilde{\bar
\alpha} \left(z- z^* \right)}\ . \ee Here \be \label{R17b}
\tilde{\bar \alpha}\equiv {4 \over \phi_0\sqrt{-3\gamma}}\ . \ee
The sign $\mp$ in (\ref{R17}) corresponds to the sign in
(\ref{R16}). Varying over $z^*$, one obtains the following
equation: \be \label{R18} 0={d^2 z \over d\tau^2} + (1
-i\tilde{\bar \alpha})\left({dz \over d\tau}\right)^2
 - \left(1+i\tilde{\bar \alpha}\right)\ .
\ee Now, the Hamiltonian $H$ conjugate to $\tau$ is given by \be
\label{R19} H=\mp {1 \over \kappa^2}\sqrt{8V_0 \over 3}\int d^3 x
\left[ {dz \over d\tau}{dz^* \over d\tau} - 1 \right]\e^{v+u -
{\bar \alpha} (v-u)}\ , \ee and the Hamiltonian constraint has the
following form: \be \label{R20} {dz \over d\tau}{dz^* \over d\tau}
=1\ . \ee By defining a new variable ${\cal Z}$ as \be \label{R21}
{\cal Z}\equiv \e^{(1 - i \tilde{\bar \alpha})z}\ , \ee Eq.~
(\ref{R18}) can be rewritten as \be \label{R22} {d^2 {\cal Z}
\over d\tau^2}=\left(1 + \tilde {\bar \alpha}^2\right){\cal Z}\ .
\ee The solution of (\ref{R22}) is given by \be \label{R23} {\cal
Z}=z_+ \e^{\tau\sqrt{1 +\tilde {\bar \alpha}^2}} + z_-
\e^{-\tau\sqrt{1 + \tilde{\bar \alpha}^2}}\ , \ee with complex
constants of integration $z_\pm$. The Hamiltonian constraint
(\ref{R20}) restricts the constants to satisfy \be \label{R24} z_+
z_-^*=- z_- z_+^*\ . \ee By using three real independent
parameters $b_\pm$ and $\theta_0$, the solution of (\ref{R24}) is
given by \be \label{R25} z_+ = b_+ \e^{i\theta_0}\ \quad z_-=ib_-
\e^{-i\theta_0}\ . \ee By using $z_\pm$ in (\ref{R25}), the metric
of the spacetime has the following form: \bea \label{R26b} ds^2
&=& -{8 \over 3V_0}\left|\left\{z_+ \e^{\tau\sqrt{1 +\tilde {\bar
\alpha}^2}} + z_- \e^{-\tau\sqrt{1 + \tilde{\bar
\alpha}^2}}\right\}^{\alpha \over 1 - i\alpha}\right|^2 d\tau^2
\nn && + \left|\left\{z_+ \e^{\tau\sqrt{1 +\tilde {\bar
\alpha}^2}} + z_- \e^{-\tau\sqrt{1 + \tilde{\bar
\alpha}^2}}\right\}^{2 \over 3(1-i\alpha)}\right|^2
\sum_{i=1,2,3}\left(dx^i\right)^2 \ . \eea

When $\tau\to \infty$, from (\ref{R15b}) and (\ref{R16}), it
follows that \bea \label{R27} a &\sim& \e^{{2\tau \over 3\sqrt{ 1
+ \tilde{\bar \alpha}^2}} + {2 \over 3\left( 1 + \tilde{\bar
\alpha}^2\right)}\Re\left\{\left( 1 + i \tilde{\bar
\alpha}\right)\ln z_+ \right\}}\ , \nn t &\sim& t_+ \mp {\sqrt{1 +
\tilde{\bar \alpha}^2} \over \tilde{\bar \alpha}^2} \sqrt{8 \over
3V_0}\e^{-{\tilde{\bar \alpha}^2 \over 1 + \tilde{\bar
\alpha}^2}\Im \left\{ \left( 1 + i \tilde{\bar \alpha}\right)\ln
z_+\right\}}\e^{ - {\tilde{\bar \alpha}^2\tau \over \sqrt{ 1 +
\tilde{\bar \alpha}^2}}}\ . \eea Here $t_+$ is a constant of the
integration. Hence, $a\propto \left(\mp \left(t - t_+\right)
\right)^{-{2 \over 3\tilde{\bar \alpha}^2}}$, which tells that the
universe is accelerating since $\ddot a \propto {2 \over
3}\tilde{\bar \alpha}^2\left({2 \over 3}\tilde{\bar \alpha}^2 +
1\right) \left(\pm \left(t - t_+\right) \right)^{-{2 \over
3\tilde{\bar \alpha}^2}-2}>0$. The effective $w$ is given by \be
\label{Rw1} w=-1 - \tilde{\bar \alpha}^2<-1\ . \ee The case
corresponds to a phantom. In general, \bea \label{Raa1}
\lefteqn{\ddot a = {3V_0 \over 8}{\cal Z}^{{1 \over
3\left(1-i\tilde{\bar \alpha}\right)}
 - {i\tilde{\bar \alpha} \over 1-i\tilde{\bar \alpha}} - 2}
{\cal Z^*}^{{1 \over 3\left(1+i\tilde{\bar \alpha}\right)}
 + {i\tilde{\bar \alpha} \over 1+i\tilde{\bar \alpha}} - 2}
\left[{4\left( 1 + \tilde{\bar \alpha}^2\right) \over 9}{\cal Z}^2{\cal Z^*}^2 \right.} \\
&& \left. - {2\left(1+i\tilde{\bar \alpha}\right)\left(2-3i\tilde{\bar \alpha}\right)
\over 9\left(1-i\tilde{\bar \alpha}\right)}z_+ z_- {\cal Z^*}^2
 - {2\left(1-i\tilde{\bar \alpha}\right)\left(2+3i\tilde{\bar \alpha}\right)
\over 9\left(1+i\tilde{\bar \alpha}\right)}z_+^* z_-^* {\cal Z}^2
\right]\ .\nonumber \eea If $\left|{\cal Z}\right|$ is large, then
$\ddot a>0$. We should note that when $z_-=0$, Eq.~ (\ref{R27})
gives an exact solution corresponding to those in Sect. \ref{S2}.
In this case, Eq.~ (\ref{R27}) is valid even if $|t|$ is not
small. The solution can be regarded as an attractor. In fact, when
$t\sim t_+$, all the solutions behave as this one.

On the other hand, when $\tau\to -\infty$ one gets \bea
\label{R29} a &\sim& \e^{-{2\tau \over 3\sqrt{ 1 + \tilde{\bar
\alpha}^2}} + {2 \over 3\left( 1 + \tilde{\bar
\alpha}^2\right)}\Re\left\{\left( 1 - i \tilde{\bar
\alpha}\right)\ln z_- \right\}}\ , \nn t &\sim& t_- \pm {\sqrt{1 +
\tilde{\bar \alpha}^2} \over \tilde{\bar \alpha}^2} \sqrt{8 \over
3V_0}\e^{-{\tilde{\bar \alpha}^2 \over 1 + \tilde{\bar
\alpha}^2}\Im \left\{ ( 1 - i \tilde{\bar \alpha})\ln
z_-\right\}}\e^{ {\tilde{\bar \alpha}^2\tau \over \sqrt{ 1 +
\tilde{\bar \alpha}^2}}}\ . \eea Here $t_-$ is a constant of
integration again. Thus, $a\propto \left(\pm \left(t - t_-\right)
\right)^{-{2 \over 3\tilde{\bar \alpha}^2}}$ and, once more, the
universe is accelerating.

The solution  (\ref{R26b}) is almost given by the analytic
continuation $\bar \alpha\to i\tilde{\bar \alpha}$ of that
(\ref{Ra1}), which corresponds to the standard (non-phantom)
scalar. The behavior obtained for the solution  (\ref{R26b}) is,
however, rather different from the non-phantom case. In the case
of a non-phantom with $\gamma>0$ in (\ref{Ra1}) ---which has been
investigated in \cite{russo}--- when
$\left|{\bar\alpha}\right|<0$, the behavior in (\ref{Ra2}) or
(\ref{Ra4}) shows that there is a singularity only in the infinite
future or past, since $\tau\to \pm \infty$ corresponds to $t\to
\pm\infty$. On the other hand, when $\left|{\bar\alpha}\right|>0$,
the behavior in (\ref{Rc5}) or (\ref{Rc6}) indicates that there
might be Big Rip \cite{brett,alexei,CKM} or Big Crunch singularity
in the finite future. Even for non-phantom matter, when the strong
energy condition is applied, a finite-time future singularity may
occur\cite{john}.
In the phantom case in Eq.~ \ref{R27}) or (\ref{R29}), $\tau\to
\pm \infty$ corresponds to $t\to 0$. The singularity occurs in the
{\it finite} future or past. Thus, if the universe is expanding,
there should be a finite-time future singularity. This singularity
is nothing but the Big Rip \cite{brett,alexei,CKM}. If there is a
singularity in the past, the universe is not expanding but
shrinking. In this sense, the solution with a past singularity is
related with that with a future singularity, by reversing the
direction of time.

It is possible to relate the action (\ref{R1}) with the action
(\ref{P3}) in the Einstein frame by identifying $g_{\mu\nu}$,
$\gamma$, and $V(\phi)$ with $g_{E\mu\nu}$, ${(d-1)\alpha^2 \over
d-2}+{\gamma \over 2}$ (with $d=4$), and $\e^{-{2\alpha \over
d-2}\phi}V(\phi)$ (with $d=4$ again), respectively. The physical
metric is obtained by rescaling the metric as the reverse of
(\ref{P2}). Instead of (\ref{R2}), we now assume that the physical
metric is  given by \be \label{R30} ds^2 = \e^{\bar\beta
\phi}\left(-dt^2 + a(t)^2 \sum_{i=1,2,3}\left(dx^i\right)^2
\right)\ , \ee For $\tau\to \pm \infty$, $\phi$ behaves as \be
\label{R31} \phi\sim \mp {4 \over \tilde{\bar
\alpha}\sqrt{-3\gamma}}\ln \left|t - t_\pm\right|\ . \ee Since $a$
behaves as $a\sim \left|t-t_\pm\right|^{- {2 \over 3\tilde{\bar
\alpha}^2}}$, if \be \label{R32} \bar\beta = - {1 \over
\tilde{\bar \alpha}}\sqrt{-{\gamma \over 3}}\ , \ee the
singularity corresponding to $\tau\to + \infty$ is cancelled but
there remains a singularity corresponding to $\tau\to -\infty$. On
the other hand, if \be \label{R33} \bar\beta = {1 \over
\tilde{\bar \alpha}}\sqrt{-{\gamma \over 3}}\ , \ee the
singularity corresponding to $\tau\to - \infty$ is cancelled but
there remains a singularity corresponding to $\tau\to +\infty$. In
the metric (\ref{R30}), the cosmological time $\tilde t$ is
defined by $d\tilde t = \pm\e^{{\beta \over 2}\phi}dt$, then in
case of (\ref{R32}), we find $\tilde t \propto \tilde t_0+ \left|t
- t_\pm\right|^{{2 \over 3\tilde{\bar \alpha}^2} + 1}$. Here
$\tilde t_0$ is a constant of  integration. The limit $t\to t_\pm$
($\tau\to \pm\infty$) corresponds to $\tilde t\to \tilde t_0$. On
the other hand, in the case of (\ref{R33}), $\tilde t \propto
\tilde t_0'+ \left|t - t_\pm\right|^{-{2 \over 3\tilde{\bar
\alpha}^2} + 1}$. Here $\tilde t_0'$ is a constant of integration.
If $\tilde{\bar \alpha}^2>{3 \over 2}$, $t\to t_\pm$ ($\tau\to
\pm\infty$) corresponds to $\tilde t\to \tilde t_0$, again. If
$\tilde{\bar \alpha}^2>{3 \over 2}$, however, $t\to t_\pm$
($\tau\to \pm\infty$) corresponds to $\tilde t\to \pm\infty$.
Hence, the singularity does not occur within finite time. This
example shows that the the type (or even the presence itself) of
the singularity is also related with the choice of physical metric
(frame).

\section{Quantum effects may change the finite time future singularity.}

Let us again start from the scalar-tensor theory with a single
scalar which can be an effective phantom: \be \label{S1} L={1
\over \kappa^2}\left(R + {\tilde{\gamma} \over
2}g^{\mu\nu}\partial_\mu\phi
\partial_\nu\phi - V(\phi)\right)\ , \ee where $\tilde{\gamma}=\pm 1$.
 It would be interesting to
investigate the quantum properties of such scalar-tensor gravity.
Indeed, it is known that the phantom theory develops a
catastrophic instability at the quantum level. Hence the point is
that taking into account quantum gravity effects (or, simply quantum
effects) could improve the
situation.

The calculation of the one-loop effective action in the former,
non-renormalizable theory may indeed be performed (using the above
parametrization and some choice for the gauge condition). The
result is \bea \label{SS2} \lefteqn{W_{\rm 1-loop}=-{1 \over 2}\ln
{L^2 \over \mu^2}\int d^4x \sqrt{-g}\left\{ {5 \over 2}V^2 -
\tilde{\gamma}\left(V'\right)^2 + {1 \over 2}\left(V''\right)^2 \right.}
\nn && + \left[{\tilde{\gamma} \over 2}V -
2V''\right]\phi_{,\mu}\phi^{,\mu}
 - \left[ {13 \over 3}V + {\tilde{\gamma} \over 12}V''\right]R
+ {43 \over 60}R_{\alpha\beta}^2 \nn
&& \left. + {1 \over 40}R^2
 - {\tilde{\gamma} \over 6}R\phi_{,\mu}\phi^{,\mu} + {5 \over 4}\left(\phi_{,\mu}\phi^{,\mu}\right)^2
\right\} \ . \eea The above one-loop action is found in Ref.
\cite{BKK}. In order to consider this effective action as a finite
quantum correction to the classical one, the cut-off $L$ should be
identified with the corresponding physical quantity. For instance,
when the universe is in the (almost) deSitter phase, the natural
choice is $L^2=\left|R\right|$, as the curvature is strong enough
and constant \cite{BO,BO1}. At the same time, in the region where
$\left|V\right|\gg \left|R\right|$, $L^2$ should be identified
with $|V|$.

Hence, even in the situation with $V=0$, starting from the action
(\ref{S1}) with the usual scalar, the phantom terms may be
induced. This happens if the universe goes through a region with
negative curvature. With account to the potential, for some
fine-tuning of $V$ one can again arrive to the QG-induced phantom
theory, which subsequently can change the universe evolution.

Here, we consider the action where $L^2$ is replaced with
$\left|R\right|$ as a simple example: \bea \label{SS3}
\lefteqn{W_{\rm 1-loop}=-{1 \over 2}\int d^4x \sqrt{-g}\ln
{\left|R\right| \over \mu^2}\left\{ {5 \over 2}V^2 -
\tilde{\gamma}\left(V'\right)^2 + {1 \over 2}\left(V''\right)^2 \right.}
\nn && + \left[{\tilde{\gamma} \over 2}V -
2V''\right]\phi_{,\mu}\phi^{,\mu}
 - \left[ {13 \over 3}V + {\tilde{\gamma} \over 12}V''\right]R
+ {43 \over 60}R_{\alpha\beta}^2 \nn
&& \left. + {1 \over 40}R^2
 - {\tilde{\gamma} \over 6}R\phi_{,\mu}\phi^{,\mu} + {5 \over 4}\left(\phi_{,\mu}\phi^{,\mu}\right)^2
\right\} \ .
\eea
The variations of this action are given by
\bea
\label{SS4}
\lefteqn{{1 \over \sqrt{-g}}{\delta W_{\rm 1-loop} \over \delta\phi}
= -{1 \over 2}\ln {\left|R\right| \over \mu^2}\left\{\left\{
{5 \over 2}V^2 - \tilde{\gamma}\left(V'\right)^2 + {1 \over 2}\left(V''\right)^2 \right\}'\right.} \nn
&& + \left[{\tilde{\gamma} \over 2}V - 2V''\right]'\phi_{,\mu}\phi^{,\mu}
 - 2 \nabla_\mu\left\{\left[{\tilde{\gamma} \over 2}V - 2V''\right]\phi^{,\mu}\right\}
 - \left[ {13 \over 3}V + {\tilde{\gamma} \over 12}V''\right]'R \nn
&& \left. + {\tilde{\gamma} \over 3}\nabla_\mu\left\{R\phi^{,\mu}\right\}
 - 5 \nabla_\mu\left\{\left(\phi_{,\rho}\phi^{,\rho}\right)\phi^{,\mu}\right\}
\right\} \ , \\
\label{SS5}
\lefteqn{{1 \over \sqrt{-g}}{\delta W_{\rm 1-loop} \over \delta g_{\mu\nu}}
=-{1 \over 2}\ln {\left|R\right| \over \mu^2}\left[{1 \over 2}g^{\mu\nu}\left\{
{5 \over 2}V^2 - \tilde{\gamma}\left(V'\right)^2 + {1 \over 2}\left(V''\right)^2 \right.\right.} \nn
&& + \left[{\tilde{\gamma} \over 2}V - 2V''\right]\phi_{,\mu}\phi^{,\mu}
 - \left[ {13 \over 3}V + {\tilde{\gamma} \over 12}V''\right]R
+ {43 \over 60}R_{\alpha\beta}^2 \nn
&& \left. + {1 \over 40}R^2
 - {\tilde{\gamma} \over 6}R\phi_{,\mu}\phi^{,\mu} + {5 \over 4}\left(\phi_{,\mu}\phi^{,\mu}\right)^2
\right\} \nn
&& - \left[{\tilde{\gamma} \over 2}V - 2V''\right]\phi^{,\mu}\phi^{,\nu}
+ \left[ {13 \over 3}V + {\tilde{\gamma} \over 12}V''\right]R^{\mu\nu} \nn
&& - \left(\nabla^\mu \nabla^\nu - g^{\mu\nu}\nabla^2\right)
\left[ {13 \over 3}V + {\tilde{\gamma} \over 12}V''\right]
 - {43 \over 30}R^\mu_{\ \rho}R^{\nu\rho} \nn
&& + {43 \over 60}\left\{\left(\nabla_\alpha\nabla^\nu R^{\alpha\mu}
+ \nabla_\alpha \nabla^\mu R^{\alpha\nu}\right) - \nabla^2 R^{\mu\nu}
 - g^{\mu\nu}\nabla_\rho \nabla_\sigma R^{\rho\sigma}\right\} \nn
&& + {1 \over 20}R R^{\mu\nu} + {1 \over 20}\left(\nabla^\mu \nabla^\nu - g^{\mu\nu}
\nabla^2 \right)R + {\tilde{\gamma} \over 6}R^{\mu\nu}\phi_{,\rho}\phi^{,\rho} \nn
&& - {\tilde{\gamma} \over 6}\left(\nabla^\mu \nabla^\nu - g^{\mu\nu}\nabla^2\right)
\left(\phi_{,\rho}\phi^{,\rho}\right) + {\tilde{\gamma} \over 6}R \partial^\mu\phi \partial^\nu\phi
 - {5 \over 2} \phi_{,\rho}\phi^{,\rho} \phi^{,\mu} \phi^{,\nu} \nn
&& + \left( - R^{\mu\nu} + \nabla^\mu \nabla^\nu - g^{\mu\nu}\nabla^2\right)\left[
 - {1 \over 2R} \left\{
{5 \over 2}V^2 - \tilde{\gamma}\left(V'\right)^2 + {1 \over 2}\left(V''\right)^2 \right. \right. \nn
&& + \left[{\tilde{\gamma} \over 2}V - 2V''\right]\phi_{,\rho}\phi^{,\rho}
 - \left[ {13 \over 3}V + {\tilde{\gamma} \over 12}V''\right]R
+ {43 \over 60}R_{\alpha\beta}^2 \nn
&& \left.\left. + {1 \over 40}R^2
 - {\tilde{\gamma} \over 6}R\phi_{,\rho}\phi^{,\rho} + {5 \over 4}\left(\phi_{,\rho}\phi^{,\rho}\right)^2
\right\} \right] \ . \eea In the case of occurrence of the Big Rip
singularity, the curvature quickly grows. However, this means that
quantum effects (e.g. quantum gravity effects) become important
not only for the early universe but also for the future universe.
These quantum effects may even become dominant when the universe
approaches the Big Rip. Suppose that the quantum correction
becomes dominant owing to the fact that $W_{\rm 1-loop}$ contains
higher derivative terms. In this case one can neglect the
classical terms. To simplify the situation even more, we assume
that the curvature and the scalar field $\phi$ are constant \be
\label{SS6} R_{\mu\nu}={3 \over l^2}g_{\mu\nu}\ ,\quad R={12 \over
l^2}\ ,\quad \phi=c\ . \ee The potential $V(\phi)$ is chosen as
the exponential function of $\phi$: \be \label{SS7} V(\phi)=V_0
\e^{-2{\phi \over \phi_0}}\ . \ee Then from (\ref{SS4}) and
(\ref{SS5}), we obtain \bea \label{SS8} 0&=&{1 \over
\sqrt{-g}}{\delta W_{\rm 1-loop} \over \delta\phi} \nn &=& -{1
\over 2}\ln {\left|R\right| \over \mu^2}\left[ -{4 \over \phi_0}
\left({5 \over 2} - {4\tilde{\gamma} \over \phi_0^2} + {8 \over
\phi_0^4}\right) V_0^2 \e^{-{4c \over \phi_0}} \right. \nn &&
\left. + {2 \over \phi_0}\left({13 \over 3}
+ {\tilde{\gamma} \over 3\phi_0^2}\right)V_0 \e^{-{2c \over \phi_0}}{12 \over l^2}\right]\ . \\
\label{SS9} 0&=&{1 \over \sqrt{-g}}{\delta W_{\rm 1-loop} \over
\delta g_{\mu\nu}} \nn &=&g^{\mu\nu}\left[ - {1 \over
4}\ln\left({12 \over l^2\mu^2}\right)\left\{ \left({5 \over 2} -
{4\tilde{\gamma} \over \phi_0^2} + {8 \over \phi_0^4}\right) V_0^2
\e^{-{4c \over \phi_0}} \right.\right. \nn && \left. - \left({13
\over 3} + {\tilde{\gamma} \over 3\phi_0^2}\right)V_0 \e^{-{2c \over
\phi_0}}{12 \over l^2} + {147 \over 5l^4}\right\} + {1 \over
8}\left({5 \over 2} - {4\tilde{\gamma} \over \phi_0^2} + {8 \over
\phi_0^4}\right)V_0^2 \e^{-{4c \over \phi_0}}\nn && \left. + {3
\over 2l^2}\left({13 \over 3} + {\tilde{\gamma} \over 3\phi_0^2}\right)V_0
\e^{-{2c \over \phi_0}} - {441 \over 40l^4}\right]\ . \eea Eq.~
(\ref{SS8}) can be solved with respect to $l^2$: \be \label{SS10}
R={12 \over l^2}=2 \left({5 \over 2} - {4\tilde{\gamma} \over \phi_0^2} +
{8 \over \phi_0^4}\right) \left({13 \over 3} + {\tilde{\gamma} \over
3\phi_0^2}\right)^{-1}V_0 \e^{-{2c \over \phi_0}}\ . \ee We should
note, however, that Eq.~ (\ref{SS9}) is not consistent with the
expression in (\ref{SS10}) in general. Then, Eq.~ (\ref{SS9})
might be regarded as an equation determining $\mu$. We should also
note that the r.h.s. in (\ref{SS10}) is not always positive. In
the case $\tilde{\gamma}>0$, when $\tilde{\gamma}^2<5$, the r.h.s. in (\ref{SS10})
is positive, but when $\tilde{\gamma}^2>5$, it is positive if $\phi_0^2 >
{4 \over 5}\left(\tilde{\gamma} + \sqrt{\tilde{\gamma}^2 - 5}\right)$ or $\phi_0^2
< {4 \over 5}\left(\tilde{\gamma} - \sqrt{\tilde{\gamma}^2 - 5}\right)$. On the
other hand, in a phantom case $\tilde{\gamma}<0$, the r.h.s. in
(\ref{SS10}) is positive if $\phi_0^2 > - {\tilde{\gamma} \over 13}$.
Anyway, there may occur  a (asymptotically) de Sitter  solution.
Thus, before entering  the Big Rip singularity, the universe
becomes a quantum de Sitter space. This qualitative discussion
indicates that the finite time future singularity may never occur
(or, at least may become milder) under the conjecture that quantum
effects become dominant just before the Big Rip. Due to the sharp
increase of the curvature invariants near the Big Rip, such a
conjecture looks quite natural.

A similar phenomenon occurs even without quantum gravity.
Indeed, let us consider again the phantom theory of Sect. 3 with the
same potential.
For the FRW background,
if we assume that $\phi$ only depends on time, the equation of motion for
$\phi$ is given by
\be
\label{dSP3}
0=-\gamma \left(\frac{d^2 \phi}{dt^2} + 3H \frac{d\phi}{dt}\right) - V'(\phi)\ .
\ee
The energy density $\rho_\phi$ is 
\be
\label{dSP4}
\rho_\phi={\gamma \over 2}\left(\frac{d\phi}{dt}\right)^2 + V(\phi)\ ,
\ee
and the FRW equation has the following form:
\be
\label{QQQ1}
{6 \over \kappa^2}H^2=\rho_\phi\ .
\ee
Here $H={\dot a \over a}$.
Then a solution of (\ref{dSP3}) and (\ref{QQQ1}) is
\be
\label{QQQ2}
\phi=\phi_0\ln \left|\frac{t_s - t}{t_1}\right|\ ,\quad
H=-\frac{\gamma \kappa^2}{4\left(t_s - t\right)}\ .
\ee
Here $t_s$ is a constant of integration and $t_1$ is given by
\be
\label{QQQ3}
t_1^2 = -\frac{\gamma \phi_0^2 \left(1 - \frac{3\gamma \kappa^2}{4}
\right)}{2V_0}\ .
\ee
Eq.(\ref{QQQ2}) shows
\be
\label{QQQ4}
a=a_0 \left|\frac{t_s - t}{t_1}\right|^{-\frac{\gamma \kappa^2}{4}}\ .
\ee
For a phantom with $\gamma<0$, $a$ grows up to infinity at $t=t_s$, which is
the Big Rip singularity \cite{brett,alexei,CKM}.

In general, Eq.(\ref{dSP3}) shows that
\be
\label{dSP5}
\frac{d\rho_\phi}{dt}= -3\gamma H \left(\frac{d\phi}{dt}\right)^2\ ,
\ee
which is positive if $\gamma<0$, $H>0$, and $\dot\phi\neq 0$. Hence, for
the phantom
with negative $\gamma$, the energy density increases in general.
We should also note that since the contribution to $\rho_\phi$ from the kinetic
 term is negative if $\gamma<0$, we find
\be
\label{BR2}
\rho_\phi \leq V(\phi)\ .
\ee
For the case $V(\phi)=0$, the energy density $\rho_\phi$ is not positive.
In general, the Big Rip singularity occurs due to the rapid increase of the
 energy density
of the phantom scalar. When $V(\phi)=0$, the singularity does not occur.
To be concrete, we also consider here matter to be dust, whose energy density is 
given by
\be
\label{BR3}
\rho_d={\rho_0 \over a^3}\ ,
\ee
with a constant $\rho_0$, which we assume to be positive.
When $V(\phi)=0$, the solution of (\ref{dSP3}) is given by
\be
\label{QBR6}
\frac{d\phi}{dt} = {c \over a^3}\ .
\ee
Here $c$ is a constant.
Then the FRW equation has the form:
\be
\label{BR3b}
{6 \over \kappa^2}H^2={\gamma c^2 \over 2a^6} + {\rho_0 \over a^3}\ .
\ee
This equation (\ref{BR3b}) can be solved easily as
\be
\label{BR4}
a^3=-{\gamma c^2 \over 2\rho_0} + {9\kappa^2 \over 4\rho_0^2}\left(t-t_s\right)\ .
\ee
Here $t_s$ is a constant of the integration.
Then, there is no singularity and there is not acceleration, either.

The Big Rip singularity in (\ref{QQQ4}) occurs since the potential 
is unbounded and goes to positive infinity when $\phi\to -\infty$.
Eq.(\ref{BR2}) tells that if $V(\phi)$ is bounded from  above and has a maximum
$V_m$ as for $V(\phi)=0$ case, the energy density does not grow up infinitely
and the Big Rip singularity does not occur.
We now assume for the large negative $\phi$, the potential
approaches  a constant:
\be
\label{QBR4}
V(\phi)\to V_m\ \left(\mbox{constant}\right)\ \mbox{when}\ \phi\to -\infty\ .
\ee
In the region, Eq.(\ref{dSP3}) reduces to
\be
\label{QBR5}
0=-\gamma \left(\frac{d^2\phi}{dt^2} + 3H \frac{d\phi}{dt}\right) \ ,
\ee
which can be solved as in (\ref{QBR6}). Then the energy density $\rho_\phi$ 
(\ref{dSP4}) has the following form:
\be
\label{QBR7}
\rho_\phi={\gamma c^2 \over 2a^6} + V_m\ .
\ee
The FRW equation becomes
\be
\label{BR5}
{6 \over \kappa^2}H^2 = {\gamma c^2 \over 2a^6} + V_m \ .
\ee
The first term in the r.h.s. could be neglected for a large universe.
Then, for large $a$, one gets the deSitter space as a solution
\be
\label{QBR11}
H^2\to {V_m \over 6\kappa^2}\ .
\ee
Thus, one way to avoid the singularity might be that, in the present
universe,  for
large negative $\phi$, there is an upper bound in the potential.

Another way to argue is to take into
 account quantum effects, say, for conformally-invariant matter.
Then the contributions coming from the conformal anomaly to the energy density
$\rho_A$ and pressure $p_A$ are (see \cite{NOOfrw})
\bea
\label{hhrA3}
\rho_A&=&- 6 b'H^4 - \left({2 \over 3}b + b''\right)
\left\{ -6 H \frac{d^2 H}{dt^2} - 18 H^2 \frac{dH}{dt}
+ 3 \left(\frac{dH}{dt}\right)^2 \right\} \\
\label{hhrAA1}
p_A&=&b'\left\{ 6 H^4 + 8H^2 \frac{dH}{dt} \right\} \nn
&& + \left({2 \over 3}b + b''\right)\left\{ -2\frac{d^3 H}{dt^3} -12
H \frac{d^2 H}{dt^2} - 18 H^2 \frac{dH}{dt}  - 9 \left(\frac{dH}{dt}\right)^2
 \right\}\ .
\eea
In general, with $N$ scalar, $N_{1/2}$ spinor, $N_1$ vector fields, $N_2$ 
($=0$ or $1$) gravitons and $N_{\rm HD}$ higher derivative conformal scalars,
 $b$, $b'$ and $b''$ are given by
\bea
\label{bs}
&& b={N +6N_{1/2}+12N_1 + 611 N_2 - 8N_{\rm HD} \over 120(4\pi)^2}\nn
&& b'=-{N+11N_{1/2}+62N_1 + 1411 N_2 -28 N_{\rm HD} \over 360(4\pi)^2}\ ,
 \quad b''=0\ .
\eea

Near the Big Rip singularity, the scale factor $a$ blows up, as in (\ref{QQQ4}),
 at $t=t_s$.
Then the curvatures behave as $R\propto \left|t - t_s\right|^{-2}$ and
they become large.
Since the quantum correction includes the square of the curvatures, the
correction becomes
large and important near the Big Rip singularity. Now the FRW equation has the 
following form
\be
\label{QBR1}
{6 \over \kappa^2}H^2 ={\gamma \over 2}\left({d\phi \over dt}\right)^2 + V(\phi)
 + \rho_A\ .
\ee
We now write $H$ and $\phi$ as
\be
\label{QQ4}
H=h_0 + \delta h \ ,\quad \phi=\phi_0\ln \left|\frac{t_s - t}{t_1}\right| + 
\delta\phi\ .
\ee
Here $h_0$, $t_s$, and $t_1$ are constants. We assume that when $t\to t_s$,
$\delta h$ and $\delta\phi$ become very small compared with the first terms,
respectively. In $H$, however, as the first term is a constant, only the
second term contributes to $\frac{dH}{dt}$ and $\frac{d^2H}{dt^2}$. From 
(\ref{dSP3}) with the same exponential potential, one obtains
\bea
\label{QQ5}
0&=&-\gamma\left( - \frac{\phi_0}{\left(t_s - t\right)^2} - \frac{3h_0}{t_s - t}\right)
+ \frac{2V_0t_1^2}{\phi_0\left(t_s - t\right)^2}\left(1 - \frac{2}{\phi_0}\delta\phi\right) \nn
&& + o\left(\left(t_s - t\right)^{-1}\right)\ ,
\eea
which gives
\be
\label{QQ6}
V_0t_1^2= - {\gamma\phi_0^2 \over 2}\ ,\quad \delta\phi=-\frac{3}{2}\left(t_s - t\right)\ .
\ee
Then from (\ref{QBR1}), we have
\be
\label{QQ7}
0=\frac{3\gamma h_0 \phi_0}{t_s - t} + 6h_0 \left(\frac{2}{3}b + b''\right)
\frac{d^2 \delta h}{dt^2} + o\left(\left(t_s - t\right)^{-1}\right)\ ,
\ee
and
\be
\label{QQ8}
\delta h=\frac{\gamma\phi_0}{2 \left(\frac{2}{3}b + b''\right)}\left(t_s - t\right)
\ln \left|\frac{t_s - t}{t_2}\right| \ .
\ee
Here $t_2$ is a constant of integration.
Since $H=\frac{\dot a}{a}$, the scale factor is
\be
a=a_0\left|\frac{t_s - t}{t_2}\right|^{\frac{\gamma\phi_0}{4 \left(\frac{2}{3}b 
+ b''\right)}
\left(t_s - t\right)^2}\e^{-h_0\left(t_s - t\right)
 - \frac{\gamma\phi_0}{8 \left(\frac{2}{3}b + b''\right)}\left(t_s - t\right)^2
+ o\left(\left(t_s - t\right)^2\right)}\ .
\ee
In $\frac{d^2 a}{dt^2}$ or $\frac{dH}{dt}$, there appears a logarithmic 
singularity. The behavior of the singularity, however, becomes rather milder 
than the case without quantum correction, where due to the singularity at 
$t=t_s$, the universe cannot develop beyond the singularity. Since the 
singularity becomes mild due to the quantum correction and $a$ and $H$ are 
finite at $t=t_s$, the universe can develop to the region $t>t_s$.
Then, essentially the Big Rip singularity is removed due to the quantum correction.


\section{Gravity assisted dark energy dominance and effective
phantom/quintessence cosmology.}

Recently, an effective phantom/quintessence description of the
late time universe was obtained via the introduction of a new
higher-derivative coupling between matter and gravity \cite{SNSO}.
It was shown that such a model may explain the gravity assisted
dark energy dominance. In the present section we will consider a
simple (scalar) example of such kind of model when standard matter
is also included.

The starting action is: \be \label{LR1} S=\int d^4 x
\sqrt{-g}\left\{{1 \over \kappa^2}R + R^\alpha L_d  + L_m\right\}
\ . \ee Here $L_d$ is the matter-like Lagrangian density (dark
energy) and $L_m$ the Lagrangian density of the (standard) matter.
By variation over $g_{\mu\nu}$, the equation of motion follows:
\be \label{LR2} 0= {1 \over \sqrt{-g}}{\delta S \over \delta
g_{\mu\nu}} = {1 \over \kappa^2}\left\{{1 \over 2}g^{\mu\nu}R -
R^{\mu\nu}\right\} + \tilde T^{\mu\nu} + T_m^{\mu\nu}\ . \ee Here
the effective energy momentum tensor $\tilde T_{\mu\nu}$ is
defined by \be \label{w5} \tilde T^{\mu\nu}\equiv  - \alpha
R^{\alpha - 1} R^{\mu\nu} L_d + \alpha\left(\nabla^\mu \nabla^\nu
- g^{\mu\nu}\nabla^2 \right)\left(R^{\alpha -1 } L_d\right) +
R^\alpha T^{\mu\nu}\ . \ee and $T^{\mu\nu}$ is given by \be
\label{LR3} T^{\mu\nu}\equiv {1 \over \sqrt{-g}}{\delta \over
\delta g_{\mu\nu}} \left(\int d^4x\sqrt{-g} L_d\right)\ . \ee The
standard matter part of the energy momentum tensor $T_m^{\mu\nu}$
is also defined as \be \label{LR3m} T_m^{\mu\nu}\equiv {1 \over
\sqrt{-g}}{\delta \over \delta g_{\mu\nu}} \left(\int
d^4x\sqrt{-g} L_m\right)\ . \ee For simplicity,  the Lagrangian
density of a free massless scalar is considered as $L_d$: \be
\label{LR4} L_d = - {1 \over 2}\partial_\mu \phi \partial_\nu
\phi\ , \ee Note that for the above $L_d$ choice and with a higher
derivative scalar curvature term in the gravitational sector (also
without standard matter),  model of this kind was discussed, with
different purposes, in Refs.\cite{dolgov}.

The equation of motion has the following form: \be \label{LR5} 0=
{1 \over \sqrt{-g}}{\delta S \over \delta \phi}={1
\over\sqrt{-g}}\partial_\mu \left(R^\alpha \sqrt{-g}
g^{\mu\nu}\partial_\nu \phi \right)\ . \ee The metric is again
chosen to describe a FRW universe with flat 3-space: \be
\label{LR6} ds^2 = - dt^2 + a(t)^2 \sum_{i=1,2,3}
\left(dx^i\right)^2\ . \ee

If one assumes that $\phi$ depends only on $t$
$\left(\phi=\phi(t)\right)$, the solution of the scalar field
equation (\ref{LR5}) is given by \be \label{LR9} \dot \phi = q
a^{-3} R^{-\alpha}\ . \ee Here $q$ is a constant of integration.
Hence \be \label{LR10} R^\alpha L_d = {q^2 \over 2 a^6 R^\alpha}\
, \ee which becomes dominant when $R$ is small (large) compared
with the Einstein term ${1 \over \kappa^2}R$ if $\alpha>-1$
$\left(\alpha <-1\right)$. Thus, one arrives at the remarkable
possibility that dark energy grows to asymptotic dominance over
the usual matter with decrease of the curvature.

Combining (\ref{LR1}) and (\ref{LR2}), one gets
$S\sim \int d^4 x \sqrt{-g}\left\{{1 \over \kappa^2}R
+ {q^2 \over 2 a^6 R^\alpha} \right\}$,
which may indicate $R\sim a^{-{6 \over \alpha + 1}}$.
Then the curvature $R$ might be stabilized to have a non-trivial minimum due to
the second term in (\ref{LR1}).

Substituting (\ref{LR9}) into (\ref{LR2}), the $(\mu,\nu)=(t,t)$
component of the equation of motion has the following form: \bea
\label{LR11} 0&=&-{3 \over \kappa^2}H^2 + \rho_d + \rho_m \nn
\rho_d&\equiv& {36q^2 \over a^6}\left(6\dot H + 12
H^2\right)^{-\alpha - 2} \left\{ {\alpha (\alpha + 1) \over
4}\ddot H H + {\alpha + 1 \over 4}{\dot H}^2 \right. \nn && \left.
+ \left(1 + {13 \over 4}\alpha + \alpha^2\right)\dot H H^2 +
\left(1 + {7 \over 2}\alpha\right) H^4 \right\} \ . \eea Here
$\rho_m$ is the energy density of the standard matter.
Specifically, when $\alpha = -1$, Eq.~ (\ref{LR11}) looks like:
\be \label{LR12} 0=-\left({3 \over \kappa^2} + {15 q^2 \over
2a^6}\right)H^2 + \rho_m\ . \ee If $\rho_m=0$, this equation has
only the trivial solution $H=0$ ($a$ is a constant).

When $\rho_m=0$, we can easily find the accelerating solution of
(\ref{LR11})\cite{SNSO}: \be \label{LR13} a=a_0 t^{\alpha + 1
\over 3}\quad \left(H={\alpha + 1 \over 3t}\right)\ , a_0^6 \equiv
{\kappa^2 q^2 \left(2\alpha - 1\right)\left(\alpha - 1\right)
\over 3\left(\alpha + 1\right)^{\alpha + 1} \left({2 \over
3}\left(2\alpha - 1\right)\right)^{\alpha + 2}}\ . \ee
Eq.~(\ref{LR13}) tells that the universe accelerates, that is,
$\ddot a>0$ if $\alpha>2$. Even for $\alpha<-1$, by changing the
time variable by $t\to t_0 - t$ ($t_0$ is a constant), the
universe is expanding and accelerating. In this case, however,
there is a Big Rip singularity at $t=t_0$.

For the matter satisfying the relation $p=w\rho$, where $p$ is the
pressure and $\rho$ the energy density, from the usual FRW
equation, one has \be \label{w1} a\propto t^{2 \over 3(w+1)} . \ee
For $a\propto t^{h_0}$ it follows that \be \label{w2} w=-1 + {2
\over 3h_0}\ , \ee and an accelerating expansion ($h_0>1$) of the
universe occurs if \be \label{w3} -1<w<-{1 \over 3}\ . \ee For the
case of (\ref{LR13}), one finds \be \label{w4} w={1 - \alpha \over
1 + \alpha}\ . \ee Then if $\alpha<-1$,  $w<-1$, what corresponds
to an effective phantom. In this case,  changing $t$ as $t_0 - t$
in (\ref{LR13}), there appears a Big Rip singularity at $t=t_0$.
In \cite{WH}, however, it has been shown that the phantom energy
with $w<-1$ makes the radius of the wormhole spacetime (when it
does occur) to increase in time and thus before the Big Rip the
radius becomes infinite and, as a result, the Big Rip singularity
may be avoided.

It is interesting to investigate the stability of the solution in
(\ref{LR13}). For this purpose, we write the scale factor $a$ as
\be \label{p1} a=a_0 t^{\alpha + 1 \over 3}\left(1 +
\delta\right)\ ,\quad \left(\left|\delta\right|\ll 1\right)\ . \ee
Here $a_0$ is given in (\ref{LR13}). From (\ref{LR11}), it follows
that\bea \label{p2} 0&=& - {2\left(2\alpha -1\right)\left(\alpha +
1\right)\left(\alpha - 1\right) \over t^4}\delta
 - {2\left(2\alpha^3 - 18 \alpha^2 - 33 \alpha -7 \right) \over t^3}{d\delta \over dt} \nn
&& + {9\alpha\left(\alpha + 4\right) \over t^2}{d^2 \delta \over
dt^2} + {9\alpha \over t}{d^3 \delta \over dt^3}\ . \eea The
solutions of (\ref{p2}) are given as: \be \label{p3} \delta
\propto t^{x_0}\ . \ee Here $x_0$ is a constant. Then
Eq.~(\ref{p2}) gives \bea \label{p3b} 0&=& 9\alpha x_0^3 +
9\alpha\left(\alpha + 1\right)x_0^2 + \left(-4\alpha^3 + 27
\alpha^2 + 48 \alpha + 14\right)x_0 \nn && -
2\left(\alpha+1\right)\left(2\alpha -1\right)\left(\alpha -
1\right)\ . \eea As clear from (\ref{w4}), when $\alpha\to
+\infty$,  $w\gtrsim -1$, which corresponds to quintessence and,
when $\alpha\to -\infty$, we find $w\lesssim -1$, which
corresponds to a phantom. Imagine that parameter $\alpha\to
\pm\infty$. If we assume $x_0 = {\cal O}(1)$, Eq.~(\ref{p3b})
reduces to \be \label{p4} 0=-4\alpha^3\left(1+x_0\right) + {\cal
O}\left(\alpha^2\right)\ . \ee Hence \be \label{p5} x_0\sim -1\ .
\ee With $x_0={\cal O}(\alpha)$, one obtains \be \label{p6}
0=-4\alpha^3 x_0 + 9\alpha^2 x_0^2 + 9\alpha x_0^3 + {\cal
O}(\alpha^3)\ . \ee As a result \be \label{p7} x_0 \sim 0, {\alpha
\over 3}, -{4\alpha \over 3}\ . \ee Note that the first solution
$x_0\sim 0$ corresponds to the solution in (\ref{p5}) since we
have assumed $x_0={\cal O}(\alpha)$. The perturbation looks as \be
\label{p8} \delta \sim\delta_1 t^{-1} + \delta_2 t^{\alpha \over
3} + \delta_3 t^{-{4\alpha \over 3}}\ . \ee Here $\delta_{1,2,3}$
are constants. The second term in (\ref{p8}) may indicate the
instability of the solution (\ref{LR13}). When $\alpha\to
+\infty$, which corresponds to quintessence, the second term may
become dominant when $t\to \infty$. On the other hand, since the
case
 $\alpha\to -\infty$
corresponds to the phantom, we replace $t\to t_0 - t$. The second
term may become dominant near the Big Rip $t\to t_0$. This might,
however, indicate that the
 Big Rip does never occur, since the Big Rip solution is unstable. In other
words, even if the present universe equation of state parameter
looks as $w\lesssim -1$, the universe might transit to another
solution corresponding to the second term in (\ref{p8}). However,
it is difficult to find the non-perturbative behavior of the
solution corresponding to the second term in (\ref{p8}).

The $\alpha\to -1$ case which corresponds to $w\to -\infty$ can be
also considered. Let us write \be \label{p9} \alpha =
-\left(1+\epsilon\right)\ , \quad \left(\epsilon\ll 1\right)\ .
\ee It is natural to assume that $\epsilon$ is positive.
 Eq.~(\ref{p3b}) gives
\be \label{p10} 0=-9\left(1+\epsilon\right)x_0^3 + 9\epsilon x_0^2
+ \left(-3 + 18\epsilon\right)x_0 + 12\epsilon + {\cal
O}\left(\epsilon^2\right)\ . \ee Its approximate solution is \be
\label{p11} x_0=4\epsilon,\ -{3 \over 2}\epsilon \pm {i \over
\sqrt{3}} \left(1 - {7 \over 2}\epsilon\right)\ . \ee The
perturbation $\delta$ is found to be given by \bea \label{p12}
\delta &=& \delta_0 \left(t_0 - t\right)^{4\epsilon} + \left(t_0 -
t\right)^{-{3 \over 2}\epsilon}\left\{ \delta_c \cos \left( {1
\over \sqrt{3}}\left(1 - {7 \over 2}\epsilon\right)\ln \left(t_0 -
t\right)\right) \right. \nn && \left. + \delta_s \sin \left( {1
\over \sqrt{3}}\left(1 - {7 \over 2}\epsilon\right)\ln \left(t_0 -
t\right)\right)\right\} \ . \eea Here $\delta_0$, $\delta_c$, and
$\delta_s$ are constants. The first term decreases near the Big
Rip at $t=t_0$ but the other terms oscillate rapidly near the Big
Rip and the amplitude becomes large, which also shows the
instability of the (transient acceleration) solution (\ref{LR13}).
This may be quite an acceptable result, with the condition that it
lasts sufficiently enough to comply with observational data. The
way to avoid a finite time future singularity due to the
instability of the accelerating cosmology may deserve some
attention.

The case $\alpha\gtrsim 1$ corresponds to $w\lesssim 0$. With \be
\label{p13} \alpha=1 + \epsilon\ ,\quad \left(0<\epsilon \ll
1\right)\ , \ee Eq.~(\ref{p3b}) gives \be \label{p14}
0=9\left(1+\epsilon\right)x_0^3 + \left(18 + 27\epsilon\right)
x_0^2 + \left(85 + 90\epsilon\right)x  - 4\epsilon + {\cal
O}\left(\epsilon^2\right)\ . \ee Its solution is \be \label{p15}
x_0={4 \over 85}+{\cal O}\left(\epsilon^2\right),\ -1 \pm {2i
\over 9}+{\cal O}\left(\epsilon\right)\ . \ee Therefore $\delta$
is given by \be \label{p16} \delta \sim \hat\delta_0 t^{4\epsilon
\over 85} + {1 \over t}\left\{ \hat\delta_c \cos \left( {2 \over
9}\ln t\right) + \hat\delta_s \sin \left( {2 \over 9}\ln
t\right)\right\} \ . \ee Here $\hat\delta_0$, $\hat\delta_c$, and
$\hat\delta_s$ are constants, too. The first term increases with
$t$, and thus the solution  (\ref{LR13}) may be unstable again.

Even with numerical calculation for general $\alpha$, there seems to be
instabilities
in the solution (\ref{LR13}) if $w<0$ although the solution could be stable if $w>0$.
This may suggest again that the acceleration of such dark energy  universe
might
be transient.

So far the standard matter contribution has been neglected. When
$\rho_m\neq 0$, the simple assumption is that $\rho_m$ behaves as
\be \label{LR16} \rho_m = \rho_0 a^{-\beta}\ . \ee Here $\rho_0$
and $\beta$ are constant.  From Eq.~(\ref{LR12}) it follows \be
\label{LR17} t=\sqrt{3 \over \rho_0 \kappa^2}\int da a^{{\beta
\over 2} - 4}\sqrt{a^6 + {5q^2\kappa^2 \over 2}}\ . \ee

For the case of a general $\alpha$, if the dark energy density is
dominant :$\rho_d\gg \rho_m$, the solution should behave as
(\ref{LR13}). Hence, the energy density $\rho_m$ of the standard
matter evolves as \be \label{LRm1} \rho_m = \rho_0 a_0^{-\beta}
t^{-{(\alpha + 1)\beta \over 3}}. \ee On the other hand, $\rho_d$
behaves as \be \label{LRm2} \rho_d \propto t^{-2}\ . \ee Since
$\beta=4$ for the radiation and $\beta=3$ for the dust and the
acceleration of the universe occurs when $\alpha>2$, we may assume
$\beta>2$ and $\alpha>2$. As a result, \be \label{LRm3} {(\alpha +
1)\beta \over 3}>2\ . \ee When time $t$ grows, $\rho_m$ decreases
further ias compared with $\rho_d$.

The essence of a gravity-assisted dark energy dominance is clearly
seen in the example below. Let the standard matter dominate, as
compared with the dark energy $\rho_m \gg \rho_d$. Then the scale
factor is given by that of the standard FRW equation: \be
\label{LRm4} a = \left({\beta^2 \rho_0 \kappa^2 \over
12}\right)^{1 \over \beta} t^{2 \over \beta}\ . \ee The energy
density of the standard matter $\rho_m$ behaves as $\rho_m \sim
t^{-2}$. On the other hand, the dark energy behaves as \be
\label{LRm5} \rho_d \sim t^{2\alpha - {12 \over \beta}}\ . \ee
Hence if \be \label{LRm6} 2\alpha - {12 \over \beta}>-2\ , \ee the
dark energy $\rho_d$ becomes larger as time passes.
Eq.~(\ref{LRm6}) can be satisfied if $\alpha, \beta>2$ as in the
dark matter dominant case.

Let us assume that in the early universe, the standard
matter/radiation is dominant. The universe evolves according to
(\ref{LRm4}). From (\ref{LRm6}), dark energy increases with time
growth. When $\rho_d \gtrsim \rho_m$, as in the present universe,
the acceleration of the universe begins. Thus, in the future
 the accelerating universe evolves with the scale factor (\ref{LR13}).
However, that is most probably a transient acceleration. This is
not strange owing to the fact that the above effective
phantom/quintessence description is achieved by means of a higher
derivative coupling between dark matter and gravity. It is known that
higher derivative gravities (see \cite{BO1} for a review) may have
problems with unitarity (stability) when they are considered as
fundamental theories. Hence, the model under discussion should be
treated as a kind of effective matter-gravity theory. It would be
interesting to analyze some astrophysical predictions of our model
(say, rotation curves of galaxies) in the way it was recently
discussed e.g. in \cite{salvatore}.

\section{Discussion.}

In summary, (phantom) scalar-tensor cosmology with an exponential
scalar potential suggests the comely possibility of a dark energy
universe with an equation of state parameter $w$ which is negative
and very close to $-1$. Convenient choices of the theoretical
parameters may shift the explicit value of $w$  from above
to below  $-1$. Moreover, such a universe naturally admits a
(transient or eternal) acceleration phase. A very nice property of
this theory is that what appears as a phantom in one frame may
look as a standard scalar in another frame. It is demonstrated
that in the situation when a finite time future singularity is
predicted by the growing phantom energy density, the consideration
of quantum gravity effects might drastically change the future of
our universe, removing the singularity in a quite natural way. From
another side, it is also shown that a higher-derivative
gravity-matter coupling term being not the phantom may in fact provide an
effective phantom/quintessence description of the late-time
universe, suggesting the possibility of a dark energy model of a brand
new type. In this
case, gravity makes dark energy to become the (evolving) main
contribution to the total energy density ---as compared with the standard
matter/radiation, which was initially dominant--- what leads to the
apearence of the phase of transient acceleration.

Current attention to phantom models as dark energy candidates
is not driven by the
internal
consistency and/or beauty of this theory, which still contains a
number of non-fully-resolved problems, as we have  already mentioned.
 Rather, it is the lack of a good theoretical understanding of
the present universe coming from more usual theories what calls for
alternative explanations to be considered, on one hand. From
another, one sees also that the cosmic zoo structure which emerges
from these alternatives
is so rich and suitable at times, that some concepts which so
far seemed to be strange (like the idea of negative energy itself)
deserve to be investigated with care and to the end. In this
respect, even the mild indications which have been reported of
a possible phantom origin coming
from string/M-theory or on the chance (described above)
to avoid the Big Rip catastrophe
by taking properly into account the quantum
effects seem indeed very promising.

\section*{Acknowledgments}

We thank  James Lidsey, Alexei Starobinsky and Paul Townsend for fruitful
discussions. This investigation has been supported in part by the
Ministry of Education, Science, Sports and Culture of Japan under
grant n.13135208 (S.N.), by the RFBR grant 03-01-00105 and LRSS
grant 1252.2003.2 (S.D.O.), and by SEEU and DGICYT (Spain), grant
PR2004-0126  and project BFM2003-00620, respectively (E.E.).

\appendix

\section{Appendix}
In this Appendix  several remarks are made about the possible
origin of the  phantom-related models coming from the higher
dimensional theories considered in the paper. This topic was
widely investigated in the Kaluza-Klein context (in relation with
the string/M-theory), and for a recent discussion the reader is
addressed to \cite{NO1}. We start from a $4+n$-dimensional
spacetime, whose metric is given by \be \label{R8} ds^2 =
\sum_{\mu,\nu=0,1,2,3}g_{\mu\nu} dx^\mu dx^\nu + \e^{2\phi(x^\mu)}
\sum_{i,j=1}^n \tilde g_{ij} d\xi^i d\xi^j\ . \ee For simplicity,
one may assume that the metric $\tilde g_{ij}$ corresponds to an
Einstein manifold, where the Ricci tensor $\tilde R_{ij}$
constructed from $\tilde g_{ij}$ is proportional to $\tilde
g_{ij}$: $\tilde R_{ij}=k \tilde g_{ij}$. Here $k$ is a constant.
When $n=1$, $k$ always vanishes ($k=0$). When $n\geq 3$, the above
metric is given as the solution of the $n$-dimensional Euclidean
Einstein equation. When $n=2$, since the 2d Einstein equation is
trivial, in the conformal gauge the above condition for the Ricci
tensor is the Liouville equation. Under the above assumptions, the
$4+n$ dimensional Einstein action with matter field $\chi$
(bosonic sector of some higher-dimensional supergravity) can be
written as \bea \label{R10} S_{4+n}&=&{1 \over \kappa^2}\int
d^{4+n} x \sqrt{-g^{(4+n)}}\left(R^{(4+n)} -{1 \over
2}\partial_\mu \chi
\partial^\mu \chi - U(\chi) \right) \nn &=& {V_n \over
\kappa^2}\int d^4 x \sqrt{-g} \e^{n\phi}\left( R +
n(n-1)g^{\mu\nu}\partial_\mu \phi \partial_\nu \phi +
nk\e^{-2\phi} \right. \nn && \left. -{1 \over 2}\partial_\mu \chi
\partial^\mu \chi - U(\chi) \right)\ . \eea Here $V_n$ is the
volume of the $n$-dimensional manifold whose metric tensor is
given by $\tilde g_{ij}$. We should note that the kinetic energy
of $\phi$ becomes negative, as for the phantom. Rescaling the
4-dimensional metric $g_{\mu\nu}$ by $g_{\mu\nu}\to
\e^{-n\phi}g_{\mu\nu}$, the action (\ref{R10}) can be rewritten as
\bea \label{R12} S_{4+n}&=& {V_n \over \kappa^2}\int d^4 x
\sqrt{-g}\left( R
 - {n(n+2) \over 2}g^{\mu\nu}\partial_\mu \phi \partial_\nu \phi \right.\nn
&& \left. + nk\e^{-(n+2)\phi} -{1 \over 2}\partial_\mu \chi \partial^\mu \chi -
\e^{-n\phi}U(\chi)\right)\ .
\eea
Now the kinetic energy of $\phi$ is positive.
If we further rescale $\phi$ by
\be
\label{P45}
\phi=\varphi \sqrt{3 \over n(n+2)}\ ,
\ee
the four-dimensional action looks like
\bea
\label{P46}
S_{4+n}&=& {V_n \over \kappa^2}\int d^4 x \sqrt{-g}\left( R
 - {3 \over 2}g^{\mu\nu}\partial_\mu \varphi \partial_\nu \varphi \right.\nn
&& \left. + nk\e^{-\varphi\sqrt{3(n+2) \over n}}  -{1 \over 2}\partial_\mu \chi
\partial^\mu \chi - \e^{-\varphi\sqrt{3n \over n+2}}U(\chi)\right)\ .
\eea The above action belongs to the same class as (\ref{PB1}) by
identifying \be \label{P47} W(\varphi,\chi)= \e^{-\varphi\sqrt{3n
\over n+2}}U(\chi)\ . \ee Comparing the above expression with
(\ref{PB3}), one obtains \be \label{P48} \sqrt{3n \over n+2}={2 +
\eta \over \varphi_0}\ . \ee Then, if \be \label{P49} 0<\varphi_0
< 2 \sqrt{n+2 \over 3n}\ , \ee   Eq.~(\ref{PB8}) follows. If \be
\label{P50} U(\chi)=W_0 \e^{{\eta \over \chi_0}\chi} \ , \ee we
can obtain  for $w$ a value less than $-1$, and correpsondigly the
universe expands with acceleration, by a proper choice of  the
parameters.

One may consider the product compactification to be more general
than (\ref{R8}): \bea \label{R26} ds^2 &=&
\sum_{\mu,\nu=0}^{d-1}g_{\mu\nu} dx^\mu dx^\nu +
\e^{2\phi^{(1)}(x^\mu)}
 \sum_{i,j=1}^n g^{(1)}_{ij} d\xi^i d\xi^j \nn
&& + \e^{\phi^{(2)}(x^\mu)} \sum_{I,J=1}^N g^{(2)}_{IJ} d\xi^i
d\xi^j\ . \eea Since the scalar curvature $R^{(d+n+N)}$ in
$(d+n+N)$-dimensional spacetime is given by \bea \label{P51}
R^{(d+n+N)}&=& R^{(d)} + k^{(1)}\e^{-2\phi^{(1)}} -
n(n+1)\partial_\mu \phi^{(1)}
\partial^\mu \phi^{(1)} - 2n \nabla^2 \phi^{(1)} \nn
&& + k^{(2)}\e^{-2\phi^{(2)}} - N(N+1)\partial_\mu \phi^{(2)}\partial^\mu \phi^{(2)}
 - 2N \nabla^2 \phi^{(2)} \nn
&& - 2nN \partial_\mu \phi^{(1)} \partial^\mu \phi^{(2)}\ , \eea
if we start from the $(d+n+N)$-dimensional action coupled with the
scalar field $\chi$ with potential $U(\chi)$, we obtain \bea
\label{P52} S^{(d+n+N)}&=& {1 \over \kappa^2}\int d^{d+n+N}x
\sqrt{-g^{(d+n+N)}}\left[ R^{(d+n+M)}
 - {1 \over 2}\partial_\mu \chi\partial^\mu \chi - U(\chi)\right] \nn
&=& {V_n V_N \over \kappa^2}\int d^d x \sqrt{-g^{(d)}}\e^{n\phi^{(1)} + N\phi^{(2)}}
\Bigl[ R^{(d)} -n \partial_\mu \phi^{(1)}\partial^\mu \phi^{(1)} \nn
&& - N \partial_\mu \phi^{(2)}\partial^\mu \phi^{(2)}
+ \partial_\mu \left(n\phi^{(1)} + N \phi^{(2)}\right)
\partial^\mu \left(n\phi^{(1)} + N \phi^{(2)}\right) \nn
&& + k^{(1)}\e^{-2\phi^{(1)}} + k^{(2)}\e^{-2\phi^{(2)}}
 - {1 \over 2}\partial_\mu \chi\partial^\mu \chi - U(\chi)\Bigr] \ .
\eea
If $U(\chi)$ is a constant, one may regard it as the cosmological
constant.
Further rescaling the metric tensor by
\be
\label{P53}
g_{\mu\nu}\to \e^{- {2 \over d-2}\left(n\phi^{(1)} + N
\phi^{(2)}\right)}g_{\mu\nu}\ ,
\ee
we get
\bea
\label{P54}
S^{(d+n+N)}&=& {V_n V_N \over \kappa^2}\int d^d x \sqrt{-g^{(d)}}
\Bigl[ R^{(d)} -n \partial_\mu \phi^{(1)}\partial^\mu \phi^{(1)}
 - N \partial_\mu \phi^{(2)}\partial^\mu \phi^{(2)} \nn
&& - {1 \over d-2} \partial_\mu \left(n\phi^{(1)} + N \phi^{(2)}\right)
\partial^\mu \left(n\phi^{(1)} + N \phi^{(2)}\right) \nn
&& + k^{(1)}\e^{- {2 \over d-2}\left(n\phi^{(1)} + N \phi^{(2)}\right)-2\phi^{(1)}}
+ k^{(2)}\e^{- {2 \over d-2}\left(n\phi^{(1)} + N \phi^{(2)}\right)-2\phi^{(2)}} \nn
&& - {1 \over 2}\partial_\mu \chi\partial^\mu \chi
 - \e^{- {2 \over d-2}\left(n\phi^{(1)} + N \phi^{(2)}\right)}U(\chi)\Bigr] \ .
\eea In this frame, the kinetic term of the matter field $\chi$
does not directly couple to the scalar fields $\phi^{(1)}$ and
$\phi^{(2)}$. Then the Newton law is not violated in the leading
order of perturbation. The kinetic terms of the fields
$\phi^{(1)}$ and $\phi^{(2)}$ can be diagonalized by \be
\label{P55} \phi^{(1)} = y^- \phi^+ + y^+ \phi^- \ ,\quad
\phi^{(2)} = y^+ \phi^+ - y^- \phi^- \ . \ee Here \be \label{P56}
y^\pm\equiv \sqrt{{1 \over 2}\left(1\pm
{\left(d-2\right)^2\left(n-N\right)^2 \left(1 + {n+N \over
d-2}\right)^2 \over 4n^2 N^2 }\right)}\ . \ee Hence \bea
\label{P54b} S^{(d+n+N)}&=& {V_n V_N \over \kappa^2}\int d^d x
\sqrt{-g^{(d)}} \Bigl[ R^{(d)} -x^+ \partial_\mu
\phi^+\partial^\mu \phi^+
 - x^- \partial_\mu \phi^- \partial^\mu \phi^-  \nn
&& + k^{(1)}\e^{- \left\{\left({2n \over d-2} +2\right)y^- + {2N \over d-2}y^+\right\}\phi^+
 - \left\{\left({2n \over d-2} +2\right)y^+ - {2N \over d-2}y^-\right\}\phi^-} \nn
&& + k^{(2)}\e^{- \left\{\left({2N \over d-2} +2\right)y^+ + {2n \over d-2}y^-\right\}\phi^+
 - \left\{-\left({2N \over d-2} +2\right)y^- - {2n \over d-2}y^+\right\}\phi^-} \nn
&& - {1 \over 2}\partial_\mu \chi\partial^\mu \chi
 - \e^{- {2 \over d-2}\left\{\left(ny^- + Ny^+\right)\phi^+
+ \left(ny^+ - Ny^-\right) \phi^-\right\}}U(\chi)\Bigr] \ . \eea
Here, \bea \label{P55b} x^\pm &\equiv& {1 \over 2}\left(n+N + {n^2
+ N^2 \over d-2}\pm \sqrt{D}\right)>0\ ,\nn D&\equiv& \left(n+N +
{n^2 + N^2 \over d-2}\right)^2 - 4nN\left(1 + {n+N \over
d-2}\right) \nn &=& \left(n-N\right)^2 \left(1 + {n+N \over
d-2}\right)^2 + {4n^2 N^2 \over \left(d-2\right)^2} > 0\ . \eea
For simplicity,  the case $k^{(2)}=U(\chi)=\chi=0$ and $d=4$ is
considered. Comparing the action (\ref{P54b}) with (\ref{PB1}), we
may identify \be \label{P56b} \varphi\leftrightarrow \sqrt{2x^+
\over 3}\phi^+\ ,\quad \kappa\chi \leftrightarrow
\sqrt{2x^-}\phi^-\ . \ee From (\ref{PB3}), one sees that \bea
\label{P57} W_0&=&-k^{(1)}\ , \nn {2+ \eta \over
\varphi_0}&=&\left\{(n + 2)y^- + N y^+\right\}\sqrt{3 \over 2x^+}\
,\nn {\eta \over \chi_0}&=& -{(n + 2)y^+ - N y^- \over
\sqrt{2x^-}}\ . \eea Using (\ref{PB6}),  an expression for $\eta$
can be found: \be \label{P58} \eta = - {2 \over 1 +\zeta}\ ,\quad
\zeta \equiv {x^- \left\{(n + 2)y^- + N y^+\right\}^2 \over x^+
\left\{(n + 2)y^+ -  N y^-\right\}^2}\ . \ee Since $x^\pm>0$
(\ref{P55b}),  $\zeta>0$ and therefore  $\eta$ satisfies
(\ref{PB8}). Then  $w>-1$ and there is no phantom. More explicitly
$w$ is given by \be \label{P59} w = -1 + {2\left\{(n+2)y^- +  N
y^+\right\} \over  3x^+(2+\eta)}\ . \ee Numerically, with $n=2$
and $N=5$,  $w=0.00269554$; with $n=3$ and $N=4$, $w=-0.0586691$.
Although not realistic, if we choose $n=6$ and $N=31$, it follows
that $w=-0.353435$.

\end{document}